# IMPROVED Ni I log(gf) VALUES AND ABUNDANCE DETERMINATIONS IN THE PHOTOSPHERES OF THE SUN AND METAL-POOR STAR HD 84937

M. P. Wood[1], J. E. Lawler[1], C. Sneden[2,3], and J. J. Cowan[4]

[1]Department of Physics, University of Wisconsin, Madison, WI 53706; mpwood@wisc.edu, jelawler@wisc.edu

[2]Department of Astronomy and McDonald Observatory, University of Texas, Austin, TX 78712; chris@verdi.as.utexas.edu

[3]Department of Astronomy and Space Sciences, Ege University, 35100 Bornova, İzmir, Turkey

[4]Homer L. Dodge Department of Physics and Astronomy, University of Oklahoma, Norman, OK 73019; cowan@nhn.ou.edu

ABSTRACT

Atomic transition probability measurements for 371 Ni I lines in the UV through near IR are reported. Branching fractions from data recorded using a Fourier transform spectrometer and a new echelle spectrograph are combined with published radiative lifetimes to determine these transition probabilities. Generally good agreement is found in comparisons to previously reported Ni I transition probability measurements. Use of the new echelle spectrograph, independent radiometric calibration methods, and independent data analysis routines enable a reduction of systematic errors and overall improvement in transition probability uncertainty over previous measurements. The new Ni I data are applied to high resolution visible and UV spectra of the Sun and metal-poor star HD 84937 to derive new, more accurate Ni abundances. Lines covering a wide range of wavelength and excitation potential are used to search for non-LTE effects.

1. INTRODUCTION

Accurate atomic transition probabilities (e.g. NIST Atomic Spectra Database[1] and Vienna Atomic Line Database[2]) are a critical component in the determination of stellar abundances. These abundances and their trends as a function of metallicity provide valuable information regarding the nucleosynthetic history of chemical elements in the Galaxy[3]. Of particular importance are old, metal-poor stars, whose abundances provide a record of the nucleosynthetic processes in the earliest generations of stars. Studies of metal-poor stars have found unexpected trends with metallicity for relative iron (Fe)-group abundances (McWilliam et al. 1995a & 1995b, McWilliam 1997, Westin et al. 2000, Cowan et al. 2002, Sneden et al. 2003, Cayrel et al. 2004, Barklem et al. 2005, Lai et al. 2008, Bonifacio et al. 2009, Roederer 2009, Suda et al. 2011, Yong et al. 2013). The [X/Fe] trends cover up to ±1 dex over the metallicity range from solar ([Fe/H] ≡ 0) to -4 (e.g. Figure 12 of McWilliam 1997). These trends have not been reconciled with current models of supernova yields in the early Galaxy.

While it may be that models of nucleosynthesis in the early Galaxy need to be revisited, or that standard abundance derivation techniques are inadequate for low metallicity stars, the possibility remains that laboratory atomic data are contributing to the observed trends. The use of modern techniques, including branching fraction measurements from a Fourier transform spectrometer (FTS) normalized with laser-induced fluorescence (LIF) lifetime measurements, have largely eliminated 1 dex errors in experimental atomic transition probabilities. In order to obtain the most accurate abundances it is best to use lines that are weak enough in the photosphere of the star of interest to avoid saturation. In studies covering a wide range of metallicity (>2 dex), this requires using many lines covering a range of excitation potential (E.P.) and log(*gf*) values that can introduce the possibility that inaccurate atomic transition probabilities are affecting the measured abundance. In higher metallicity stars, one uses weaker lines with high E.P. values in the first spectra (neutral atoms are a minor ionization stage in stars of interest). As metallicity decreases, one must switch to stronger lines with lower E.P. values and possibly to second spectra lines (singly ionized atoms are the dominant ionization stage), assuming suitable second spectra lines exist in the wavelength region being analyzed. Unfortunately, due to the energy level structure of Ni II, the majority of lines from this species occur in the vacuum-UV and one must rely primarily on Ni I lines for abundance determinations. The strength of these high and low E.P. lines can vary by orders of magnitude, making it difficult to accurately measure both with small uncertainties. Conversely, the observed abundance trends may be caused by failure of the 1D/LTE (one-dimensional/local thermodynamic equilibrium) approximations of photospheric models, traditionally used for abundance determinations, in metal-poor stars of interest (e.g. Asplund 2005). Metal-poor giant stars are favored in these studies to provide a large photon flux for high signal-to-noise (S/N), high-resolution spectra. The low-density atmospheres of giant stars, combined with the reduced electron pressure from the lower metal content, results in lower collision rates which may lead to departures from LTE. The two explanations detailed above for the unexpected trends can be investigated by improving atomic transition probabilities for Fe-group elements. If the trends are the result of 3D/non-LTE effects, one approach is to map anomalous abundances measurements for various lines covering a range of E.P. and wavelength in a wide range of stellar types. While this is a time-consuming approach, the alternative of accurately incorporating these effects in photospheric models is also quite difficult. The lack of key atomic data, including cross sections and rate constants for inelastic collisions of H and He with metal atoms and ions, is a major challenge for non-LTE modeling (e.g. Asplund 2005). If the observed trends persist, even after targeted searches for non-LTE effects using improved laboratory data, it would be a strong indication that models of nucleosynthetic yields in the early Galaxy are inaccurate or incomplete and need to be revisited.

---

[1] Available at http://physics.nist.gov/PhysRefData/ASD/lines_form.html and http://physics.nist.gov/cgi-bin/ASBib1/Fvalbib/search_form.cgi.
[2] Available at http://www.astro.uu.se/~vald/php/vald.php.
[3] We adopt standard spectroscopic notations. For elements X and Y, the relative abundances are written [X/Y] = $\log_{10}(N_X/N_Y)_{star} - \log_{10}(N_X/N_Y)_{\odot}$. For element X, the "absolute" abundance is written log ε(X) = $\log_{10}(N_X/N_H) + 12$. Metallicity we be considered equivalent to the [Fe/H] value.

An effort is underway to reduce transition probability uncertainties of selected neutral and singly-ionized Fe-group lines. The work on Mn I and Mn II (Den Hartog et al. 2011) focuses on multiplets that either cover small wavelength ranges or are Russell Saunders (LS) multiplets, and occasionally both. Multiplets covering small wavelength ranges are less affected by systematic uncertainty in the radiometric calibration, while transition strengths for LS multiplets can be checked against theoretical calculations. Given these advantages, it is possible to reduce the log(*gf*) uncertainties to 0.02 dex with $2\sigma$ confidence. These small uncertainties are difficult to achieve and are only practical under favorable conditions. The recent work on Ti I (Lawler et al. 2013) and Ti II (Wood et al. 2013) instead takes a broader approach by attempting to measure every possible line connecting to upper levels with previously measured lifetimes. This approach results in a much larger set of lines with transition probability measurements, though often with higher uncertainties (0.02 to ~0.10 dex). However, the small (~0.02 dex) uncertainties achieved by Den Hartog et al. (2011) are not required for detecting non-LTE effects in metal-poor stars. Non-LTE effects of 0.5 to 1 dex are found in selected metal-poor stars but are confined to Mn I resonance lines connecting to the ground level (Sobeck et al. 2014, in preparation).

This work on Ni I follows the same broad approach used for Ti I and Ti II by attempting transition probability measurements for every possible line from 57 odd-parity and 9 even-parity upper levels with lifetime measurements by Bergeson and Lawler (1993). The result is a set of 371 log(*gf*) values covering a wide range of E.P. and wavelength, with uncertainties ranging from 0.02 dex for dominant branches to ~0.12 dex for weak branches widely separated in wavelength from the dominant branch(es). The uncertainties on dominant branches are primarily from the LIF lifetime measurements while the uncertainties on weak branches, often the most important for abundance determinations, result primarily from the branching fraction measurements. Systematic effects are the dominant source of uncertainty in branching fraction measurements, and the use of both a FTS and echelle spectrograph in our study serves to better quantify and control the systematic uncertainties.

In Section 2 and Section 3 we describe our laboratory data sets from the FTS and echelle spectrograph, in Section 4 we discuss the derivation of Ni I branching fractions from these data, and in Section 5 we present the new transition probabilities with comparison to previous laboratory results. Finally, in Section 6 and Section 7 we apply the new Ni I data to determine the photospheric Ni abundances of the Sun and metal-poor star HD 84937, using many lines covering a range of E.P., wavelength, and log(*gf*) values to search for non-LTE effects.

## 2. FOURIER TRANSFORM SPECTROMETER DATA

As in much of our previous branching fraction work, this Ni I branching fraction study makes use of archived FTS data from both the 1.0 m FTS previously at the National Solar Observatory (NSO) on Kitt Peak and the Chelsea Instruments FT500 UV FTS at Lund University in Sweden. The NSO 1.0 m FTS has a large etendue (like all interferometric spectrometers), a resolution limit as small as 0.01 cm$^{-1}$, wavenumber accuracy to 1 part in 10$^8$ broad spectral coverage from the near UV to IR, and a high data collection rate (Brault 1976).

Unfortunately the NSO FTS has been dismantled, and while there are plans to restore it to full operational status at a university laboratory, it is currently unavailable to guest observers. The Chelsea Instruments FT500 has a resolution limit as small as 0.025 cm$^{-1}$, wavenumber accuracy to better than 1 part in 10$^7$, and coverage through the UV down to 1700 Å (Thorne et al. 1987). Table 1 lists the 37 FTS spectra used in our Ni I branching fraction study. All NSO spectra, raw interferograms, and header files are available in the NSO electronic archives.[4]

Multiple FTS spectra are needed to determine high-quality branching fractions. Optimum sensitivity is achieved for different spectral ranges using various beam splitter, filter, and detector combinations. Lamps with a Ne buffer gas allow for the correction of blends between Ni and Ar lines. In addition, spectra are needed with lamps operating at a range of current. Overlapping visible-UV and IR spectra of the lamps operating at high currents are needed for high S/N measurements on very weak branches to all known lower levels. Conversely, one also needs visible-UV spectra of the lamps operating at low currents in which the dominant branches are optically thin. Lines of Ni I are particularly prone to optical depth problems, even at the lowest currents available in the listed FTS spectra. The unusually severe optical depth problems are likely due to some combination of: (1) the relatively simple energy level structure of this atom which yields reduced Boltzmann factor dilution of lower level populations, (2) narrow line profiles from a near absence of hyperfine structure and small isotope shifts, (3) short upper level lifetimes which yield exceptionally strong lines to ground and low metastable levels, and (4) a high sputtering rate in the hollow cathode discharge (HCD) lamps. The optical depth concerns are addressed using the echelle spectrograph described in Section 3.

A relative radiometric calibration of the FTS is essential for the measurement of accurate emission branching fractions. As in our past branching fraction studies we make use of the Ar I and Ar II line calibration technique. Sets of well-known branching ratios for Ar I and Ar II lines have been established for this purpose in the 4300 - 35000 cm$^{-1}$ range by Adams & Whaling (1981), Danzmann & Kock (1982), Hashiguchi & Hasikuni (1985), and Whaling et al. (1993). Intensities for these lines are measured and compared to the known branching ratios in order to construct a relative radiometric calibration. The technique is internal to the HCD lamps and captures the wavelength-dependent response of the detectors, spectrometer optics, lamp windows, and any reflections which contribute to the measured signal. We can extend the calibration beyond 35000 cm$^{-1}$ by comparing measured intensities of Ni I lines in the FTS spectra to the intensities of the same lines measured with the echelle spectrograph. As described in the next section, the relative radiometric calibration of the echelle spectrograph is established using a NIST-traceable deuterium (D$_2$) lamp, which is routinely checked against a NIST-calibrated Ar mini-arc and provides reliable UV calibration to 50000 cm$^{-1}$.

---

[4] Available at http://nsokp.nso.edu/.

## 3. ECHELLE SPECTROGRAPH DATA

As mentioned in the previous section, a major motivation for the construction of a powerful new echelle spectrograph at the University of Wisconsin is the closure of the NSO 1.0 m FTS. A further motivation is the need to reduce optical depth errors as a source of systematic uncertainty in branching fraction measurements, especially on weak lines. Ni is especially prone to optical depth errors and it is therefore essential to supplement the FTS data with echelle data in order to provide the most accurate atomic transition probabilities. FTS instruments suffer from multiplex noise in which quantum statistical (Poisson) noise from all spectral features, in particular the strong visible and near-IR branches, is smoothly redistributed throughout the entire spectrum. This can be a hindrance when measuring weak transitions, which are often the most important for accurate Fe-group abundance determinations.  Often, as the lamp current is reduced, the weak lines become comparable to the multiplex noise before the dominant branch(es) from the common upper level are optically thin.  A dispersive spectrograph is free from multiplex noise and can provide adequate S/N on these astrophyscially important weak lines even at very low lamp current, reducing the possibility of optical depth errors.

A brief overview of the echelle spectrograph is given here.  The instrument is a 3.0 m focal length grating spectrograph incorporating a large (128×254 mm ruled area), coarse (23.2 grooves mm$^{-1}$) echelle grating blazed at 63.5°.  Attached to the grating spectrograph is a custom 0.5 m focal length order separator.  The order separator is rotated into a plane orthogonal to the grating spectrograph, using a prism to separate the many overlapping high orders from the echelle grating in the transverse direction and imaging them onto a large UV-sensitive CCD (Princeton Instruments PI-SX:2048, 2048×2048 13.5 μm pixels).  The orthogonal rotation serves to compensate for the inherent astigmatism of the 3.0 m grating spectrograph, allowing the entrance slits to be replaced with precision machined pinholes.   The echelle spectrograph has a resolving power of 250,000 with a 50 μm entrance pinhole, broad wavelength coverage with a few CCD frames, and superb UV sensitivity.  While the instrument has the advantage of being free from multiplex noise, it does suffer from reduced resolution and wavenumber precision compared to a FTS.  The complete design and performance of the echelle spectrograph, including a detailed aberration analysis, is provided by Wood & Lawler (2012).

In addition to the 37 FTS spectra listed in Table 1, the 80 CCD frames of spectra from commercial Ni HCD lamps listed in Table 2 are part of this Ni I branching fraction study.  These spectra are radiometrically calibrated using a NIST-traceable D$_2$ standard lamp to provide a UV continuum.  This lamp is periodically checked against a NIST-calibrated Ar mini-arc to ensure an accurate UV calibration. The use of standard lamps to calibrate a FTS is often difficult due to ghosts, and instead we rely on the calibration method described in Section 2, but it is our preferred method for calibrating the echelle spectrograph. The use of a standard lamp also allows for branching fraction measurements to much shorter wavelengths than is achievable using Ar I and Ar II branching ratios since the lamps are calibrated at wavelengths down to 2000 Å.

## 4. Ni I BRANCHING FRACTIONS

All possible transitions wavenumbers between known energy levels from the 2013 NIST Atomic Spectra Database (Kramida et al. 2013) that satisfy both the parity change and $|\Delta J| \leq 1$ selection rules are computed and used during this branching fraction analysis. Transitions which violate these two selection rules are suppressed by a factor of $\sim 10^6$ and are typically unimportant for stellar abundance studies. These two selection rules are obeyed throughout the periodic table whereas many important Fe-group transitions violate the $\Delta S = 0$ and $|\Delta L| \leq 1$ selection rules of LS coupling. We can make measurements for branching fractions as weak as 0.0001, and therefore systematic errors from missing branches are negligible in this study. Ni has five naturally occurring isotopes, one of which ($^{61}$Ni, nuclear spin I = 3/2) has hyperfine structure due to a non-zero nuclear spin. However, this isotope has a low solar system abundance of 1.1%, and Ni I lines are rather narrow in the FTS data. As a result, component structure is neglected in this branching fraction study.

Branching fraction measurements are attempted for all 57 odd-parity and 9 even-parity upper levels with lifetime measurements by Bergeson & Lawler (1993) and are completed for 52 odd-parity and all 9 even-parity levels. We note that branching fractions for the 9 even-parity upper levels were reported earlier by Wickliffe & Lawler (1997) using a subset of the FTS data employed in this work. Levels for which branching fraction measurements could not be completed have a strong branch with a severe blending problem. As in our previous work, thousands of possible spectral line observations are analyzed in both FTS and echelle spectra to calculate the branching fractions. Integration limits and non-zero baselines are set "interactively" during data analysis. Non-zero baselines are necessary for the echelle spectra, which are not dark-signal corrected, and are occasionally needed for the FTS spectra when a line falls on the wing of a dominant feature. A simple numerical integration technique is used to determine un-calibrated Ni I line intensities because of unresolved isotopic and/or hyperfine structure that causes variations in the observed line widths. This same integration technique is also used on selected Ar I and Ar II lines to establish a relative radiometric calibration of the FTS spectra.

Branching fraction uncertainties depend on the S/N of the data, the line strengths, and the wavelength separation of lines from a common upper level. Branching fraction uncertainty always migrates to the weakest lines because branching fractions sum to 1.0 by definition. Uncertainties on weak lines near the dominant branch(es) from the common upper level tend to be limited by S/N. For lines that are widely separated in wavelength from the dominant branch(es), systematic errors in the radiometric calibration tend to be the dominant source of uncertainty. The systematic uncertainty in the calibration is estimated using the product of $0.001\%/cm^{-1}$ and the wavenumber difference between the line of interest and the dominant branch from the common upper level, as presented and tested by Wickliffe et al. (2000). The calibration uncertainty is combined with the standard deviation of measurements from multiple spectra to determine the total branching fraction uncertainty. The final uncertainty, especially for lines widely separated from the dominant branch(es) of the common upper level, is primarily systematic and it is therefore impractical to state whether these uncertainties represent 1σ or 2σ error bars. The use of data from both the FTS and echelle spectrograph, combined with independent radiometric calibration techniques, is important in assessing and controlling systematic uncertainty.

## 5. Ni I TRANSITION PROBABILITIES AND COMPARISON TO EARLIER MEASUREMENTS

Absolute transition probabilities are given for 371 lines of Ni I in Table 3. Branching fraction measurements from a combination of FTS and echelle data are normalized with published radiative lifetimes (Bergeson & Lawler 1993). Air wavelengths in the table are computed using the standard index of air (Peck & Reeder 1972) and Ni I energy levels from the 2013 NIST Atomic Spectra Database (Kramida et al. 2013).

Often lines must be omitted if they are too weak to have reliable S/N, have uncertain classifications, or are too seriously blended to be separated. The effect of these problems lines can be seen by summing all transition probabilities for a given upper level in Table 3 and comparing the sum to the inverse upper level lifetime (Bergeson & Lawler 1993). The sum is typically > 90% of the inverse level lifetime. While these problem lines have large fractional uncertainty in their branching fractions, this does not have a significant effect on the uncertainties of the lines kept in Table 3. The transition probabilities uncertainties quoted in Table 3 are found by combining branching fraction uncertainties and radiative lifetime uncertainties in quadrature.

Figure 1 compares log(*gf*) values from Wickliffe & Lawler (1997) to results of this study as a function of wavelength for 76 lines in common. The horizontal line represents perfect agreement and the error bars are from this work only. Figure 2 shows the same 76 lines plotted as a function of the log(*gf*) measured in this study. These lines from high-lying even-parity upper levels are relatively easy to measure because they are not affected by optical depth errors and each multiplet is confined to a relatively small wavelength region. Wickliffe & Lawler use a subset of the FTS data used in this study for branching measurements and use the same lifetime measurements by Bergeson & Lawler (1993) for an absolute normalization, and as such, the agreement is very good. The use of an expanded set of FTS measurements in our work allows for a reduction in transition probability uncertainties compared to the Wickliffe & Lawler measurements and also facilitates the measurement of additional lines.

Figure 3 compares log(*gf*) values as a function of wavelength for 70 lines in common between this work and the measurements of Blackwell et al. (1988). The horizontal line indicates perfect agreement and the error bars represent uncertainties from this study only. Blackwell et al. claim exceptional relative accuracy (about ±0.7%) for their absorption data, but claim only around ±7% absolute accuracy. We have added a correction of +0.024 dex (about 5%) to their results in the figure in order to clarify the relative differences. Figure 4 shows the same 70 lines in common plotted as a function of the log(*gf*) measured in this study. The outlier at 3362.8 Å is both well centered and has reliable S/N in our data and we detect no problem with our measured value.

Figures 5-8 are comparisons of our measured Ni I transition probabilities to the 2013 NIST Atomic Spectra Database (Kramida et al. 2013). The database assigns an accuracy grade to each transition probability, and for the wavelength range covered in this study, there are no Ni

I lines in the database with accuracy grades of B (≤ 10%) or better. Figure 5 compares transition probabilities versus wavelength for 131 lines in common with accuracy grades of C+ (≤ 18%) and C (≤ 25%). Individual error bars represent uncertainties from this work, and the dotted lines represent ±25% differences in *f*-values. The central line represents perfect agreement at a logarithmic difference of 0. There is generally good agreement, with >80% of lines agreeing within combined uncertainties. Figure 6 shows the same 131 lines plotted as a function of the log(*gf*) measured in this work, with the lines and error bars having the same meaning as in Figure 5.

Figure 7 compares 100 lines in common between this work and the 2013 NIST Atomic Spectra Database with accuracy grades of D+ (≤ 40%) and D (≤ 50%) as a function of wavelength. The central line and the error bars have the same meanings as in Figure 5. The dotted lines in Figure 7 represent ±50% differences in *f*-values. Again there is good agreement between this work and the database, with 84% of lines agreeing within combined uncertainties. Figure 8 shows the same 100 lines plotted in Figure 7 as a function of the log(*gf*) measured in this work, with the lines and error bars having the same meaning as in Figure 7.

## 6. THE NICKEL ABUNDANCE IN THE SOLAR PHOTOSPHERE

We apply our new Ni I transition probability data to re-determine the Ni abundance of the solar photosphere. Our analytical procedure is nearly identical to that used in Lawler et al. (2013) and Wood et al. (2013). The main difference with our past papers is that we include isotopic line shifts for Ni I lines when such data are available (see the Appendix) and they appreciably affect total line broadening.

As in previous papers by our group, we first compute relative absorption strengths of Ni I lines as

$$\text{STR} \equiv \log(gf) - \theta\chi$$

with the $\log(gf)$ given in Table 3, excitation energies $\chi$ (eV), and inverse temperature $\theta = 5040/T$ (we assume $\theta = 1.0$ for this rough calculation). The STR values are plotted as a function of wavelength in Figure 9. Red circles call attention to those lines that we use in the solar abundance computations. These relative strengths apply only to Ni I transitions and cannot be compared to those of Ni II or to species of other elements because they do not include elemental abundances and Saha ionization factors. However, to a good approximation, the line strengths for a single species depend on the transition probabilities and the Boltzmann excitation factors that make up the STR factors. In Figure 9 we place a horizontal line at STR = –5.75. Examination of several solar photospheric Ni I lines (see below) suggests that lines of this strength will have reduced widths $\log(RW) = \log(EW/\lambda) \sim -6$, approaching the weak-line limit of features that are useful in a solar abundance analysis. Only 14 of our Ni I lines have STR $< -5.75$. Like similar STR plots in our previous papers, the strongest lines are at shorter wavelengths; for Ni I they occur exclusively below 4000 Å.

Since almost all of the Ni I transitions in this study should be stronger than the weak-line limit, we consider the entire set of 371 lines for the determination of a new solar Ni abundance. We first visually inspect each line in the electronic version[5] of the solar center-of-disk spectrum (Delbouille et al. 1973), and consult the Moore et al. (1966) solar line identification compendium. This procedure results in elminination of a few transitions whose photospheric line strengths are undetectably weak, and many more that are significantly compromised (especially for $\lambda < 4000$ Å) by blending with various atomic and/or molecular species. We are left with nearly 100 Ni I lines deserving of more careful study.

---

[5] http://bass2000.obspm.fr/solar_spect.php

We compute synthetic spectra for these surviving transitions and compare them to the Delbouille et al. (1973) solar spectrum. The synthetic spectra are convolved with Gaussian broadening functions to reproduce the combined effects of the very narrow spectrograph instrumental profile (resolving power, $R = \lambda/\Delta\lambda > 5\times10^5$) and solar macroturbulence. Spectrum line list construction is described by Lawler et al. (2013) in detail. Atomic and molecular lines in small spectral regions surrounding individual Ni I lines are assembled from the Kurucz (2011) database.[6] The transition probabilities are updated with recent lab results, including the Ti I and Ti II data from Lawler et al. (2013) and Wood et al. (2013), and adjustments are made to log($gf$) values of lines with no lab data in order to best reproduce the solar spectrum. We input these lists and the Holweger & Müller (1974) empirical model photosphere to the current version of the LTE line analysis code MOOG[7] (Sneden 1973). The comparison of observed and synthetic spectra result in the elimination of more lines due to unacceptably large blending of the Ni I lines by other transitions.

The final set of 76 lines are then subjected to a full synthetic spectrum analysis, this time with the inclusion of isotopic substructure in the computations when available from lab studies (see the Appendix). Isotopic fractions of the five stable Ni isotopes are (Chang 2013)[8]: f($^{58}$Ni) = 0.6808, f($^{60}$Ni) = 0.2622, f($^{61}$Ni) = 0.0114, f($^{62}$Ni) = 0.0363, and f($^{64}$Ni) = 0.0093. The three heaviest isotopes account for only 5.7% of the total solar-system Ni content. For $^{58}$Ni and $^{60}$Ni wavelengths we use the data given in the Appendix. However, there are far fewer lab data on shifts of the minor isotopes than on shifts of $^{58}$Ni and $^{60}$Ni. Fortunately, Ni I isotopic line shifts depend almost linearly on nuclear mass. Therefore in our syntheses we approximate $^{61,62,64}$Ni as a single isotope and assign it an isotopic fraction of 0.057. For a given line, we assume that the wavelength of this minor isotope is shifted from λ($^{60}$Ni) by the same amount and in the same direction that λ($^{60}$Ni) is shifted from λ($^{58}$Ni).

---

[6] http://kurucz.harvard.edu/linelists.html
[7] Available at http://www.as.utexas.edu/~chris/moog.html
[8] http://atom.kaeri.re.kr/

In Figure 10 we show sample synthetic spectra for Ni I 7414.5 Å and 7714.3 Å. These were chosen for display because they have some of the largest isotopic splitting of any of the lines considered in the present work. The larger widths of the total line profiles compared to those of the individual isotopic components can be easily seen in the figure. The main effect of the inclusion of isotopic subcomponents is to desaturate the transition. This leads to smaller derived abundances for strong Ni I lines with respect to those that are derived with single-line approximations. The magnitude of the decrease varies with isotopic splitting and line strength, but ranges from 0.01–0.08 dex for the lines of this study. Note that isotopic shifts are much smaller for lines in the blue-UV spectral domain. We ignored isotopic effects for lines with λ(58Ni) – λ(60Ni) < 0.01Å; the abundance changes for such lines are too small to be detected in our analysis.

The abundances from individual lines are listed in Table 4, in which we also record line wavelengths, excitation energies, oscillator strengths, and notes on whether isotopic substructure is included in the synthetic spectra computations. The line abundances are plotted as functions of wavelengths, excitation energies, and transition probabilities in Figure 11. There are no obvious trends beyond a small offset for the lines in the E.P. = 1.5–2.0 eV range compared with other Ni I lines. There are 33 of the 75 transitions in the E.P. = 1.5–2.0 eV range and all have wavelengths in the visible or near IR spectral region, and are all quite weak, with branching fractions between 0.003 and 0.0001 (mean ~0.001). The less crowded spectral region of these lines and their lack of saturation makes them very convenient in abundance measurements. Unfortunately, their log(*gf*) measurements are exceptionally difficult. The dominant branches from the upper levels of these 33 transitions are in the UV. This requires a relative radiometric calibration over a large wavenumber range and, as always, uncertainty migrates to the weak branches. Furthermore, it is necessary to bootstrap line ratio measurements through a sequence of line pairs in spectra recorded with lower and lower currents in order to measure branching fractions for these weak lines with good S/N and without optical depth errors. The resulting transition probabilities of these 33 lines have uncertainties of ~ 15% (~ 0.065 dex). This is near the "edge" of our lab measurement capabilities.

Abundance values from the new Ni I data suggest a possible small non-LTE effect in the solar photosphere for Ni I lines with E.P. = 1.5–2.0 eV. However, this effect is only one to two times larger than the uncertainties on the transition probabilities, and there are some additional uncertainties from the abundance determinations. Thus our data may hint at, but certainly do not demonstrate, a small non-LTE effect in solar photospheric Ni I line formation. To our knowledge, non-LTE modeling has not been extensively explored. A few studies have specifically considered the 6769.64 Å intercombination line, which is important for helioseismolgy (e.g. Bruls 1993). Recently, Vieytes & Fontenla (2013) have provided an improved Ni I atomic model that can aid non-LTE investigations. Such studies should be pursued in the future.

From the total set of 76 lines we derive a new solar photospheric Ni abundance: <log ε(Ni)> = 6.277 ± 0.004 with σ = 0.055. The mean abundance from Ni I is only slightly larger than the recommended solar photospheric abundances of 6.22 ± 0.04 (Asplund et al. 2009) and

6.23 ± 0.04 (Lodders et al. 2009).  As in our past studies, our mean abundance incorporates all 76 lines without any weighting.

## 7.  THE NICKEL ABUNDANCE OF METAL-POOR STAR HD 84937

Fe-group ($21 \leq Z \leq 30$) abundances in low-metallicity stars hold vital clues to early Galactic nucleosynthesis from massive stars.  Theory can often predict detailed abundance distributions of these elements (e.g. Kobayashi et al. 2006).  However, most abundance studies of these elements in actual metal-poor stars may not be reliable because:  (a) the neutral species of Fe-group elements are most easily detected in visible ($\lambda > 4000$Å) spectra, yet the vast majority of these elements exist in the singly-ionized state; and (b) there are concerns about the reliability of standard analytical assumptions (e.g. LTE, plane-parallel geometry) used in deriving abundances.  These concerns need to be attacked on many fronts.  The work of our group, first on rare-earth elements and now the Fe-group, is concentrating on increasing the quality and quantity of basic transition probability data.

The strongest lines of all Fe-group species lie in the UV spectral region, but this wavelength domain is very line-rich even in metal-poor giant stars.  Therefore, as in our studies of Ti I and Ti II transition probabilities (Lawler et al. 2013, Wood et al. 2013), we concentrate on exploring the spectrum of HD 84937, a metal-poor main-sequence turnoff star ($T_{eff}$ = 6300 K, log g = 4.0,  [Fe/H] = –2.15, and $v_t$ = 1.5 km s$^{-1}$).  While lines in the visible spectral region of HD 84937 are usually very weak, they are stronger and more numerous in the UV.  Additionally, the number of ionized-species transitions is greatly increased, allowing tests of the Saha ionization balance in several elements.

We derive the Ni abundance in HD 84937 in the same manner as in our previous Ti I and Ti II studies. See Lawler et al. (2013) for a detailed description of the optical *ESO VLT UVES* and *HST/STIS* UV high-resolution spectra. These combined spectra yield continuous coverage from 2300 Å to 6800 Å. We select viable Ni I transitions in the HD 84937 spectrum and derive line-by-line abundances as done in Lawler et al. (2013). In these computations we neglect isotopic splitting for two reasons. First, all lines of Ni I in the visible wavelength region ($\lambda >$ 4000 Å) are weak and thus abundances derived from them are unaffected by isotopic broadening of the line profiles. Second, although the lines are often stronger at shorter wavelengths, the isotopic wavelength splits are very small. The mean difference in the UV is $<|\lambda(58Ni) - \lambda(60Ni)|> = 0.002$ Å, undetectable with our data.

Line-by-line abundance information for the 77 Ni I lines analyzed in HD 84937 is given in Table 5, and in Figure 11 we plot the abundances as functions of their wavelengths, excitation energies, and transition probabilities. There are no obvious abundance trends with any of these quantities. In particular, the Ni I data do not show a systematic decrease in abundance in the Balmer continuum region (approximately 3100–3650 Å) that is seen in HD 84937 for Fe I by Lawler et al. (2013) and for Ti II by Wood et al. (2013). We derive a mean abundance $<\log \varepsilon(Ni)> = 3.888 \pm 0.008$ with $\sigma = 0.068$.

A few Fe-group elements (e.g., Ti and Fe) are represented by many detectable transitions of both neutral and ionized species that are accessible to ground-based high-resolution spectroscopy. Unfortunately, most Ni II lines with extant laboratory transition probability measurements are in the vacuum-UV wavelength region. Moore et al. (1966) list 23 lines of Ni II in the range 2988-4362 Å, but all of them arise from high excitation energy levels ($\chi \geq 2.9$ eV), almost all are weak and/or blended in the solar spectrum, and none have reliable log(*gf*) values.

Therefore we ignore Ni II in our solar analysis. However, the *HST/STIS* spectrum of HD 84937 exhibits many lines of this species that were subjected to a lab analysis by Fedchak & Lawler (1999). There are 19 transitions from their study in the *HST/STIS* spectral range ($\lambda > 2300$ Å). In Figure 12 we display observed and synthetic spectra of two such lines. These are chosen to also contain lines of Ni I from the present study. It is important to note that lines of Ni II are all relatively strong, and thus derived abundances from them are sensitive to our adopted microturbulence value for HD 84937. However, the Ni I lines in the *HST/STIS* spectra are also typically are strong, and so changing the microturbulence will affect both species in the same fashion, leaving abundance ratios derived from Ni I and Ni II lines relatively unchanged.

We can identify eight usable Ni II transitions in HD 84937, whose parameters and abundance estimates are given in Table 6, and we show them (along with the Ni I line results) in Figure 13. For these lines we derive $<\log \varepsilon(Ni)> = 3.89 \pm 0.04$ with $\sigma = 0.10$, in excellent agreement with the mean abundance from Ni I. Given the small number of Ni II transitions used, we cannot draw firm conclusions from the concordance between the Ni species abundances, but with the present data we do not see evidence of major departures from LTE in the ionization balance for Ni in HD 84937. Further lab studies of weaker Ni II transitions in the visible spectral region would be welcome.

Following the discussion of Lawler et al. (2013), we estimate that internal line-to-line scatter uncertainties are $\leq 0.04$ dex. For external error estimates, we derive abundances of typical Ni I and Ni II lines with model atmosphere parameters varied in accord with the HD 84937 uncertainties. For Ni I we obtain: $\Delta(\log \varepsilon) \approx +0.12$ for $\Delta(T_{eff}) = +150$K; $\Delta(\log \varepsilon) \approx -0.02$ for $\Delta(\log g) = +0.3$; $\Delta(\log \varepsilon) \approx -0.01$ for $\Delta([M/H]) = -0.3$; and $\Delta(\log \varepsilon) \approx +0.00$ to $+0.15$ for $\Delta(v_t) = -0.25$, depending on whether the lines are on the linear or saturated part of the curve-of-growth. The responses of Ni I line abundances to temperature and gravity changes are understandable from Saha ionization balances of both Ni I and H⁻, and they are essentially identical to those of Fe I because the Ni and Fe ionization energies differ by only 0.3 eV. Thus, to the extent that LTE conditions apply to the line formation of these two elements, [Ni/Fe] has little dependence on the assumed $T_{eff}$ and $\log(g)$ parameters for HD 84937. Note also that the assumed microturbulent velocity does not influence the Ni abundance much since most Ni I lines in the visible spectral region are very weak.

We repeat the model parameter variation for the UV spectral region where Ni II lines are detected. There is no substantial difference from the visible to the UV in the responses of Ni I lines to model atmsopheric parameter changes. For Ni II: $\Delta(\log \varepsilon) \approx +0.06$ for $\Delta(T_{eff}) = +150K$; $\Delta(\log \varepsilon) \approx -0.09$ for $\Delta(\log g) = +0.3$; $\Delta(\log \varepsilon) \approx -0.01$ for $\Delta([M/H]) = -0.3$; and $\Delta(\log \varepsilon) \approx +0.00$ to $+0.25$ for $\Delta(v_t) = -0.25$, depending on whether the lines are on the linear or saturated part of the curve-of-growth. However, as mentioned, above both Ni I and Ni II lines in the UV have comparable strengths, muting the dependence on assumed microturbulent velocity when comparing the abundances from these two Ni species.

Here we have shown that the Ni abundance in HD 84937 is well-determined from Ni I and Ni II lines ranging from the visible to the UV spectral region. We cannot detect departures from LTE in the Ni ionization equilibrium. However, we repeat the cautions of Wood et al. (2013) and others: we have not investigated the statistical equilibrium of Ni without the restrictive assumptions of LTE and one-dimensional atmospheric models. We cannot quantify their influence on derived Ni abundances for this star, and urge future detailed consideration of such issues.

## 8. IMPLICATIONS FOR Fe-GROUP NUCLEOSYNTHESIS

Our group has been working to determine precise atomic parameters of Fe-group species in order to confront more directly the nucleosynthetic origin of these elements. Recent results include: for Mn I and Mn II, Den Hartog et al. (2011); for Ti I, Lawler et al. (2013); for Ti II, Wood et al. (2013), and Ni I (this paper, which also includes a new look at Ni II). We are using the new experimental atomic data to derive accurate abundances of these elements in metal-poor halo stars, similar to our efforts for the rare-earth elements (REE; see e.g, Den Hartog et al. 2006; Lawler et al. 2007, 2008, 2009; Sneden et al. 2009). Such well-determined experimental data for the Fe-group elements have been lacking in the past, leading to significant stellar abundance uncertainties.

The Fe-group elements are formed in core-collapse supernovae explosions (see Thielemann et al. 1996) and then ejected into space where they are incorporated into new stars, including into halo stars early in the history of the Galaxy. Thus, observations of these elements in halo stars provide insight into early Galactic star formation and element nucleosynthesis. In addition, the Fe-group elements are synthesized in a different manner (charged-particle synthesis) than the REE (neutron-capture processes), and thus the observational abundances of these two different element groups sample different regimes of the halo progenitors. Specifically, the Fe-group elements are formed in the silicon-burning region of the exploding supernovae. Some of these elements, such as Ni, are (mostly) synthesized as a result of complete silicon fusion. In other cases, such as for Ti, the temperature in the region never rises high enough, instead leading to incomplete Si fusion. The details of how this nucleosynthesis occurs depend critically upon the supernova parameters, including the location of the mass cut (the point in the star where the mass is ejected), the explosion energy, the mass of the supernova progenitor, and the neutron excess (e.g. Nakamura et al. 1999). Thus, precise stellar abundance values can constrain the conditions that occur in the nucleosynthesis zones of supernova, and thereby provide insights into how supernovae explode.

It should also be noted that while supernovae are responsible for Fe-group element synthesis, there is no clear consensus on the astrophysical site for rapid neutron-capture synthesis (the *r*-process), which is the regime where the REE are produced. Previous work has suggested neutron-star binary mergers and neutron-star winds (rather than supernovae) as the site for this neutron-capture synthesis (Lattimer & Schramm 1974,1976; Rosswog et al. 1999; Janka et al. 1999; Wanajo & Janka 2012). Thus comparisons among the stellar observations of the REE and the Fe-group elements might in fact be sampling nucleosynthesis results from two different sites. Such observational constraints and comparisons might provide further clues about the sites for the earliest heavy element synthesis.

Using the Ni abundance we find for HD 84937 of log $\varepsilon$(Ni) = 3.89 with our new solar value of log $\varepsilon$(Ni) = 6.28 yields [Ni/H] = –2.39 for this star. The Fe abundance of HD 84397 has recently been re-determined, yielding log $\varepsilon$(Fe) = 5.18 and a metallicity of [Fe/H] = –2.32 (Lawler et al. 2013). In that paper, the authors derive a relative titanium abundance of [Ti/Fe] = +0.47. Our new results yield values of [Ni/Fe] = –0.07 and [Ti/Ni] = +0.54 for HD 84937. Lawler et al. (2013) argue that the majority of supernovae models have difficulty in reproducing the observed Ti abundance in HD 84937, as seen in other metal-poor stars (e.g., Kobayashi et al. 2006). Our new precise measurement of [Ti/Ni] will provide further constraints on supernova models.[9]

---

[9] Complicating the theoretical interpretation is that while $^{56}$Ni (which decays to $^{56}$Fe) is the dominant element for complete Si fusion, this element can also be synthesized in incomplete Si fusion (see Nakamura et al. 1999).

Our new Ni abundance result for HD 84937 provides one new data point that can be placed in the context of Galactic chemical evolution. Models that predict mean abundance ratios as a function of metallicity depend sensitively on a number of parameters including the initial mass function and supernova yields. Our value of [Ni/Fe] = –0.07 can be compared to previous determinations of [Ni/Fe] as a function of [Fe/H] (see Figures 1 & 2 of Henry et al. (2010) for a compilation of various values from the literature). We show a similar observation/theory comparison in Figure 14, where we have plotted the [Ni/Fe] ratios as a function of metallicity from a number of sources including Gratton & Sneden (1991), Feltzing & Gustafsson (1998), Cayrel et al. (2004), and Barklem et al. (2005). Our new [Ni/Fe] value for HD 84937 is shown as a large green dot in Figure 14. There is a great deal of scatter at metallicities [Fe/H] < –2, some of which may be intrinsic, but some is probably the result of imprecise abundance values resulting from older and non-experimental atomic data. Superimposed on the plot are theoretical chemical evolution abundance curves from the work of Kobayashi et al. (2006) and Kobayashi et al. (2011). The plateau for these curves is based upon the chosen initial mass function, IMF, and the yields from both supernovae (SNe) and hypernovae (HNe). The agreement with our new [Ni/Fe] value for HD 84937 is very encouraging, differing by only +0.14 from these models. Nevertheless, even this small offset might serve as a further constraint on such models, suggesting slightly higher values of [Ni/Fe] resulting from the yields of core-collapse SNe and HNe at low metallicities. At higher metallicities there is a significant increase in [Ni/Fe] as a result of the increasing Ni and Fe production from Type Ia SNe. Clearly, more such precise abundance values will be required to understand early Galactic chemical evolution and constrain the various parameters (e.g. yields) and models.

## 9. SUMMARY

Accurate absolute transition probabilities are reported for 371 lines of Ni I from the UV through near IR. Branching fraction are measured from a combination of archived FTS spectra and new data recorded with an echelle spectrograph. The use of the echelle spectrograph allows for optical depth errors in Ni I to be addressed and leads to an overall reduction in systematic uncertainties. These branching fractions are combined with previously reported radiative lifetime measurements to produce the absolute transition probabilities. Generally good agreement is found in comparison to previously reported Ni I transition probabilities. The new Ni I data are applied to re-determine the Ni abundance in the photospheres of the Sun and the metal-poor star HD 84937. There is good agreement with the mean of other [Ni/Fe] observations in the metallicity regime of this star, and only a small discrepancy with theoretical predictions.

The authors acknowledge the contribution of N. Brewer on data analysis for this project. This work is supported in part by NASA grant NNX10AN93G (J.E.L.) and NSF grants AST-0908978 and AST-1211585 (C.S.). This paper was completed while C.S. was on a University of Texas Faculty Research Assignment, in residence at the Department of Astronomy and Space Sciences of Ege University. Financial support from the University of Texas and The Scientific and Technological Research Council of Turkey (TÜBITAK, project No. 112T929) are greatly appreciated. We thank Chiaki Kobayashi for providing us with her data, and thank her and Maria Bergemann and for helpful comments.

## APPENDIX

Isotopic wavelength shifts for 303 of the 371 Ni I lines with measured transition probabilities in this study are compiled in Table 7. Previously reported measurements of the isotopic wavenumber splitting by Schroeder & Mack (1961), Steudel et al. (1980), and Litzén et al. (1993) are combined and used to determine the shifted $^{58}$Ni and $^{60}$Ni energy levels using a least squares technique. All measured isotopic splits between the upper and lower Ni I levels appearing in Table 3 are included, with equal weights given to the three studies listed above. The shifted energy levels are used to calculate the shifted isotopic wavelengths in Table 7 using the standard index of air (Peck and Reeder 1972). Due to a lack of available laboratory measurements, wavelength shifts for the minor isotopes $^{61,62,64}$Ni are not included in Table 7. These isotopes together account for only 5.7% of the solar system Ni content. In cases where isotope shift data are used in abundance determinations, we approximate the $^{61,62,64}$Ni isotopes as a single "isotope" and assign it an isotopic fraction of 0.057. Since Ni isotopic wavelength shifts depend almost linearly on isotope mass, the wavelength for this combined "isotope" is shifted from $\lambda(^{60}$Ni$)$ by the same amount and in the same direction that $\lambda(^{60}$Ni$)$ is shifted with respect to $\lambda(^{58}$Ni$)$. However, since this is only a rough approximation, the wavelengths for this combined "isotope" are not included in Table 7.

FIGURE CAPTIONS

Figure 1. Comparison of log(*gf*)s from Wickliffe & Lawler (1997) to results of this study as a function of wavelength for 76 lines in common. The solid horizontal line indicates perfect agreement. Wickliffe & Lawler use a subset of the FTS data used in this work for branching fraction measurements and use the same lifetime measurements by Bergeson & Lawler (1993) for an absolute normalization. The log(*gf*) difference are computed from the Einstein A coefficients to reveal small differences that are obscured by truncating the log(*gf*) values to 0.01 dex. The error bars are from this study only.

Figure 2. Comparison of log(*gf*)s from Wickliffe & Lawler (1997) to results of this study as a function of log(*gf*) from this work. See Figure 1 for additional details.

Figure 3. Comparison of log(*gf*)s from Blackwell et al. (1993) to results of this study as a function of wavelength for 70 lines in common. The solid horizontal line indicates perfect agreement. Blackwell et al. claim exceptional relative accuracy (about ±0.7%) for their absorption data, but claim only about ±7% absolute accuracy. We have added a correction of +0.024 dex (about 5%) to their results in the plot in order to clarify relative differences. The error bars are from this work only.

Figure 4. Comparison of log(*gf*)s from Blackwell et al. (1993) to results of this study as a function of log(*gf*) from this work. See Figure 3 for additional details.

Figure 5.  Comparison of log(*gf*)s with accuracy rank C (≤25%) and C+ (≤18%) from the online NIST Atomic Spectra Database by Kramida et al. 2013 (also in text by Fuhr et al. 1988) to results of this study as a function of wavelength for 131 lines in common.  The solid horizontal line indicates perfect agreement and the dashed lines indicate ±25% differences.  The error bars are from this work only.

Figure 6.  Comparison of log(*gf*)s for the same 131 lines shown in Figure 5, this time plotted as a function of log(*gf*) from this study.  See Figure 5 for additional details.

Figure 7.  Comparison of log(gf)s with accuracy rank D (≤50%) and D+(≤40%) from the online NIST Atomic Spectra Database by Kramida et al. 2013 (also in text by Fuhr et al. 1988) to results of this study as a function of wavelength for 100 lines in common.  The solid horizontal line indicates perfect agreement and the dashed lines indicate ±50% differences.  The error bars are from this work only.

Figure 8.  Comparison of log(*gf*)s for the same 100 lines shown in Figure 8, this time plotted as a function of log(*gf*) from this study.  See Figure 7 for additional details.

Figure 9.  Relative strengths of Ni I lines of this study, plotted as a function of their wavelengths; see text for definitions of terms.  The vertical blue line denotes the atmospheric cutoff wavelength. The horizontal blue line denotes the approximate strength of barely detectable Ni I lines (reduced widths log(*RW*) = –6).  Red circles indicate lines employed in our solar photospheric abundance analysis.

Figure 10. Synthetic and observed photospheric spectra of two Ni I lines that have significant isotopic substructure. The green open circles represent every 6th point from the Delbouille et al. (1973) solar center-of-disk spectrum. The lines representing syntheses of individual isotopic components and the combined synthesis are identified in the legend in panel (a). See the text for more information on the $^{61,62,64}$Ni component.

Figure 11. Solar photospheric abundances of Ni I lines plotted as functions of wavelength (panel a), lower excitation energy (panel b), and oscillator strength (panel c). The solid horizontal line represents the mean abundance, and the two dotted lines are placed ±1σ from the mean.

Figure 12. Observed and synthetic spectra of two small wavelength intervals in the UV spectrum of HD 84937 that contain both Ni I and Ni II lines used in this study. The observed *HST/STIS* spectrum is shown with open circles. The synthetic spectra have been computed for abundances that are color-coded as given in the legend of panel (b). The synthesis labeled "–∞" is computed without any Ni contribution. The line labeled "(Ni II)" in panel (a) was not included in the lab study of Fedchak & Lawler (1999).

Figure 13. Abundances derived from Ni I and Ni II lines in HD 84937, plotted as functions of wavelength (panel a), lower excitation energy (panel b), and oscillator strength (panel c). The solid horizontal line represents the mean abundance, and the two dotted lines are placed ±1σ from the mean.

Figure 14. [Ni/Fe] abundance ratios as a function of [Fe/H] metallicities. Observed abundances are included from four literature surveys as identified in the figure legend and the text, and our new value for HD 84937 is shown as a large green dot. The theoretical abundance predictions (blue lines) are from Kobayashi et al. (2006) and Kobayashi et al. (2011) as identified in the figure legend.

Table 1. Fourier transform spectra of Ni hollow cathode discharge (HCD) lamps. Spectra with indices 1 through 15 were recorded using the UV Chelsea Instruments FT500 at Lund University, Sweden. Spectra with indices 16 through 37 were recorded using the 1.0 m FTS on the McMath telescope at the National Solar Observatory, Kitt Peak, AZ.

| Index | Date | Serial Number | Lamp Type[a] | Buffer Gas | Lamp Current (mA) | Wavenumber Range (cm$^{-1}$) | Limit of Resolution (cm$^{-1}$) | Coadds | Beam Splitter | Filter | Detector[b] |
|---|---|---|---|---|---|---|---|---|---|---|---|
| 1 | 1994 Nov. 4 | 658 | Commercial HCD | Ar | 25 | 28436 - 56873 | 0.070 | 22 | SiO$_2$ | | R166 Solar Blind PMTs |
| 2 | 1994 Nov. 4 | 659 | Commercial HCD | Ar | 20 | 28436 - 56873 | 0.070 | 22 | SiO$_2$ | | R166 Solar Blind PMTs |
| 3 | 1994 Nov. 4 | 660 | Commercial HCD | Ar | 15 | 28436 - 56873 | 0.070 | 22 | SiO$_2$ | | R166 Solar Blind PMTs |
| 4 | 1994 Nov. 4 | 661 | Commercial HCD | Ar | 10 | 28436 - 56873 | 0.070 | 22 | SiO$_2$ | | R166 Solar Blind PMTs |
| 5 | 1994 Nov. 4 | 662 | Commercial HCD | Ar | 5 | 28436 - 56873 | 0.070 | 22 | SiO$_2$ | | R166 Solar Blind |

|    |            |     |               |       |     |               |       |    |                  |             | PMTs              |
|----|------------|-----|---------------|-------|-----|---------------|-------|----|------------------|-------------|-------------------|
| 6  | 1994 Nov. 4 | 664 | Commercial HCD | Ar    | 25  | 22117 - 44234 | 0.050 | 22 | $SiO_2$          | BG24        | 1P28 & R1516 PMTs |
| 7  | 1994 Nov. 4 | 665 | Commercial HCD | Ar    | 20  | 22117 - 44234 | 0.050 | 22 | $SiO_2$          | BG24        | 1P28 & R1516 PMTs |
| 8  | 1994 Nov. 4 | 667 | Commercial HCD | Ar    | 15  | 22117 - 44234 | 0.040 | 22 | $SiO_2$          | BG24        | 1P28 & R1516 PMTs |
| 9  | 1994 Nov. 4 | 669 | Commercial HCD | Ar    | 10  | 22117 - 44234 | 0.050 | 22 | $SiO_2$          | BG24        | 1P28 & R1516 PMTs |
| 10 | 1994 Nov. 4 | 670 | Commercial HCD | Ar    | 5   | 22117 - 44234 | 0.050 | 22 | $SiO_2$          | BG24        | 1P28 & R1516 PMTs |
| 11 | 1994 Nov. 4 | 671 | Commercial HCD | Ar    | 3   | 22117 - 44234 | 0.050 | 22 | $SiO_2$          | BG23        | 1P28 & R1516 PMTs |
| 12 | 1994 Nov. 5 | 674 | Commercial HCD | Ar    | 10  | 15798 - 31596 | 0.050 | 22 | $SiO_2$          | BG23        | 1P28 & R1516 PMTs |
| 13 | 1994 Nov. 5 | 677 | Commercial HCD | Ar    | 5   | 15798 - 31596 | 0.035 | 22 | $SiO_2$          | BG23        | 1P28 & R1516 PMTs |
| 14 | 1994 Nov. 4 | 673 | Commercial HCD | Ar    | 5   | 15798 - 31596 | 0.050 | 22 | $SiO_2$          | BG23        | 1P28 & R1516 PMTs |
| 15 | 1994 Nov. 4 | 672 | Commercial HCD | Ar    | 3   | 15798 - 31596 | 0.050 | 22 | $SiO_2$          | BG23        | 1P28 & R1516 PMTs |
| 16 | 1985 July 31 | 7  | Custom HCD    | Ne-Ar | 820 | 7181 - 46955  | 0.064 | 7  | UV               |             | Mid Range Si PDs  |
| 17 | 1985 July 31 | 9  | Custom HCD    | Ne-Ar | 80  | 7181 - 46955  | 0.064 | 7  | UV               |             | Mid Range Si PDs  |
| 18 | 1979 Dec. 29 | 1  | Custom HCD    | Ar    | 800 | 4789 - 39199  | 0.054 | 12 | UV               | CS 9-54     | Super Blue Si PDs |
| 19 | 1979 Dec. 28 | 5  | Custom HCD    | Ar    | 625 | 13607 - 37456 | 0.045 | 5  | UV               | $CuSO_4$ + CS 9-54 | Super Blue Si PDs |
| 20 | 1979 Dec. 27 | 1  | Custom HCD    | Ar    | 400 | 13607 - 37456 | 0.045 | 7  | UV               | $CuSO_4$ + CS 9-54 | Mid Range Si PDs  |

| # | Date | | | | | | | | | | |
|---|---|---|---|---|---|---|---|---|---|---|---|
| 21 | 1981 June 15 | 2 | Custom HCD | Ar | 144 | 6924 - 36533 | 0.043 | 8 | UV | WG 295 | Mid Range Si PDs |
| 22 | 1981 June 20 | 1 | Custom HCD | Ar | 580 | 8986 - 30601 | 0.037 | 20 | UV | CS 4-97 + GG 375 | Mid Range Si PDs |
| 23 | 1981 June 17 | 1 | Custom HCD | Ar | 550 | 14878 - 36386 | 0.043 | 8 | UV | $CuSO_4$ | Mid Range Si PDs |
| 24 | 1981 Sept. 22 | 1 | Custom HCD | Ar | 670 | 18529 - 36203 | 0.035 | 25 | UV | $CuSO_4$ | Super Blue Si PDs |
| 25 | 1983 Feb. 24 | 2 | Custom HCD | Ar | 2000 | 16020 - 36658 | 0.047 | 7 | UV | $CuSO_4$ | Mid Range Si PDs |
| 26 | 1983 Feb. 24 | 1 | Custom HCD | Ar | 1000 | 16020 - 36658 | 0.047 | 16 | UV | $CuSO_4$ | Mid Range Si PDs |
| 27 | 1983 Feb. 12 | 8 | Custom HCD | Ar | 80 | 15587 - 36081 | 0.044 | 8 | UV | $CuSO_4$ | Mid Range Si PDs |
| 28 | 1996 Sept. 4 | 8 | Commercial HCD | Ar | 9 | 0 - 34998 | 0.053 | 4 | UV | | Super Blue Si PDs |
| 29 | 1996 Sept. 4 | 9 | Commercial HCD | Ar | 6 | 0 - 34998 | 0.053 | 4 | UV | | Super Blue Si PDs |
| 30 | 1996 Sept. 4 | 10 | Commercial HCD | Ar | 3 | 0 - 34998 | 0.053 | 4 | UV | | Super Blue Si PDs |
| 31 | 1996 Sept. 7 | 20 | Commercial HCD | Ar | 0.8 | 0 - 34998 | 0.053 | 10 | UV | | Mid Range Si PDs |
| 32 | 1980 Mar. 20 | 2 | Custom HCD | Ar | 640 | 3571 - 9937 | 0.012 | 16 | Vis | | InSb |
| 33 | 1983 Feb. 25 | 3 | Custom HCD | Ar | 2000 | 1985 - 7811 | 0.011 | 4 | UV | C 4300 + Wedged Ge | Mid Range Si PD + InSb |
| 34 | 1983 Feb. 25 | 2 | Custom HCD | Ar | 1000 | 1985 - 7811 | 0.011 | 4 | UV | C 4300 + Wedged Ge | Mid Range Si PD + InSb |
| 35 | 1984 July 26 | 18 | Custom HCD | Ar | 1500 | 7982 - 45407 | 0.054 | 8 | UV | WG 295 | Mid Range Si PD + |

| | | | | | | | | | | | | |
|---|---|---|---|---|---|---|---|---|---|---|---|---|
| | | | | | | | | | | | | R166 Solar Blind |
| 36 | 1981 June 21 | 5 | Custom HCD | Ne | 800 | 24449 - 34522 | 0.040 | 40 | UV | $CuSO_4$ + CS 7-51 | Mid Range Si PDs | |
| 37 | 1981 June 20 | 2 | Custom HCD | Ne | 780 | 13479 - 30601 | 0.037 | 18 | UV | CS 4-96 | Mid Range Si PDs | |

[a]Lamp types include Commercial sealed HCD lamps typically used in Atomic Absorption Spectrophotometers for analytical chemistry and Custom water cooled HCD lamps for high current operation.

[b]Detector types include 1P28 Photomultiplier Tubes (PMTs), R166 Solar Blind PMTs, R1516 PMTs, Super Blue silicon (Si) Photodiodes (PDs), Mid Range Si PDs, and indium antimonide (InSb) detectors for the near infrared.

Table 2. Echelle spectra of commercial Ni HCD lamps.

| Index | Date | Serial Numbers[a] | Buffer Gas | Lamp Current (mA) | Wavelength Range (Å) | Resolving Power[b] | Coadds | Expos. Time (s) |
|---|---|---|---|---|---|---|---|---|
| 38-42 | 2012 Dec. 5 | 1, 3, 5, 7, 9 | Ne | 1 | 2200-3900 | 250,000 | 44 | 120 |
| 43-27 | 2012 Dec. 10 | 1, 3, 5, 7, 9 | Ne | 2 | 2200-3900 | 250,000 | 60 | 90 |
| 48-52 | 2012 Dec. 4 | 11, 13, 15, 17, 19 | Ne | 3 | 2200-3900 | 250,000 | 40 | 60 |
| 53-57 | 2012 Dec. 4 | 1, 3, 5, 7, 9 | Ne | 5 | 2200-3900 | 250,000 | 40 | 60 |
| 58-62 | 2012 Dec. 17 | 1, 3, 5, 7, 9 | Ar | 1 | 2200-3900 | 250,000 | 6 | 900 |
| 63-67 | 2012 Dec. 18 | 1, 3, 5, 7, 9 | Ar | 2 | 2200-3900 | 250,000 | 6 | 900 |
| 68-72 | 2012 Dec. 19 | 1, 3, 5, 7, 9 | Ar | 3 | 2200-3900 | 250,000 | 18 | 300 |
| 73-77 | 2013 Jan. 4 | 1, 3, 5, 7, 9 | Ar | 5 | 2200-3900 | 250,000 | 36 | 150 |
| 78-82 | 2013 Jan. 31 | 1, 3, 5, 7, 9 | Ne | 1 | 2200-3900 | 125,000 | 80 | 90 |
| 83-87 | 2013 Feb. 2 | 1, 3, 5, 7, 9 | Ne | 1 | 2200-3900 | 125,000 | 80 | 90 |
| 88-92 | 2013 Jan. 29 | 1, 3, 5, 7, 9 | Ar | 1 | 2200-3900 | 125,000 | 6 | 1200 |
| 93-97 | 2013 Jan. 30 | 1, 3, 5, 7, 9 | Ar | 1 | 2200-3900 | 125,000 | 6 | 1200 |
| 98-102 | 2013 Feb. 13 | 1, 3, 5, 7, 9 | Ne | 2 | 2000-2800 | 250,000 | 12 | 600 |
| 103-107 | 2013 Feb. 14 | 1, 3, 5, 7, 9 | Ne | 3 | 2000-2800 | 250,000 | 40 | 180 |
| 108-112 | 2013 Feb. 15 | 1, 3, 5, 7, 9 | Ne | 5 | 2000-2800 | 250,000 | 80 | 90 |
| 113-117 | 2013 Feb. 16 | 1, 3, 5, 7, 9 | Ar | 5 | 2000-2800 | 250,000 | 24 | 300 |

[a] At least 3 CCD frames are needed to capture a complete echelle grating order in the UV. In the above data 5 CCD frames are used to provide redundancy and a check for lamp drift.

[b] Resolving power is adjusted by changing the diameter of the precision-machined entrance pinholes from 50 μm (which gives R~250,000) to 100 μm (R~125,000).

Table 3. Experimental Atomic Transition Probabilities for 371 lines of Ni I organized by increasing wavelength in air.

| Wavelength in air[a] (Å) | Upper Level Energy[b] (cm$^{-1}$) | Parity | J | Lower Level Energy[b] (cm$^{-1}$) | Parity | J | Transition Probability (10$^6$ s$^{-1}$) | log$_{10}$(gf) |
|---|---|---|---|---|---|---|---|---|
| 2121.3903 | 47328.784 | od | 2 | 204.787 | ev | 3 | 24.6 ± 1.3 | -1.08 |
| 2125.6261 | 47030.102 | od | 3 | 0.000 | ev | 4 | 3.54 ± 0.24 | -1.77 |
| 2129.9541 | 47139.337 | od | 2 | 204.787 | ev | 3 | 3.53 ± 0.24 | -1.92 |
| 2134.9235 | 47030.102 | od | 3 | 204.787 | ev | 3 | 10.1 ± 0.7 | -1.32 |
| 2147.7838 | 47424.785 | od | 1 | 879.816 | ev | 2 | 45.4 ± 2.7 | -1.03 |

Notes. –Table 3 is available in its entirety via the link to the machine-readable version above.

[a] Wavelength values computed from energy levels using the standard index of air from Peck & Reeder 1972.

[b] Energy levels, parities, and J values are from online 2012 NIST Atomic Spectra Database by Kramida et al.

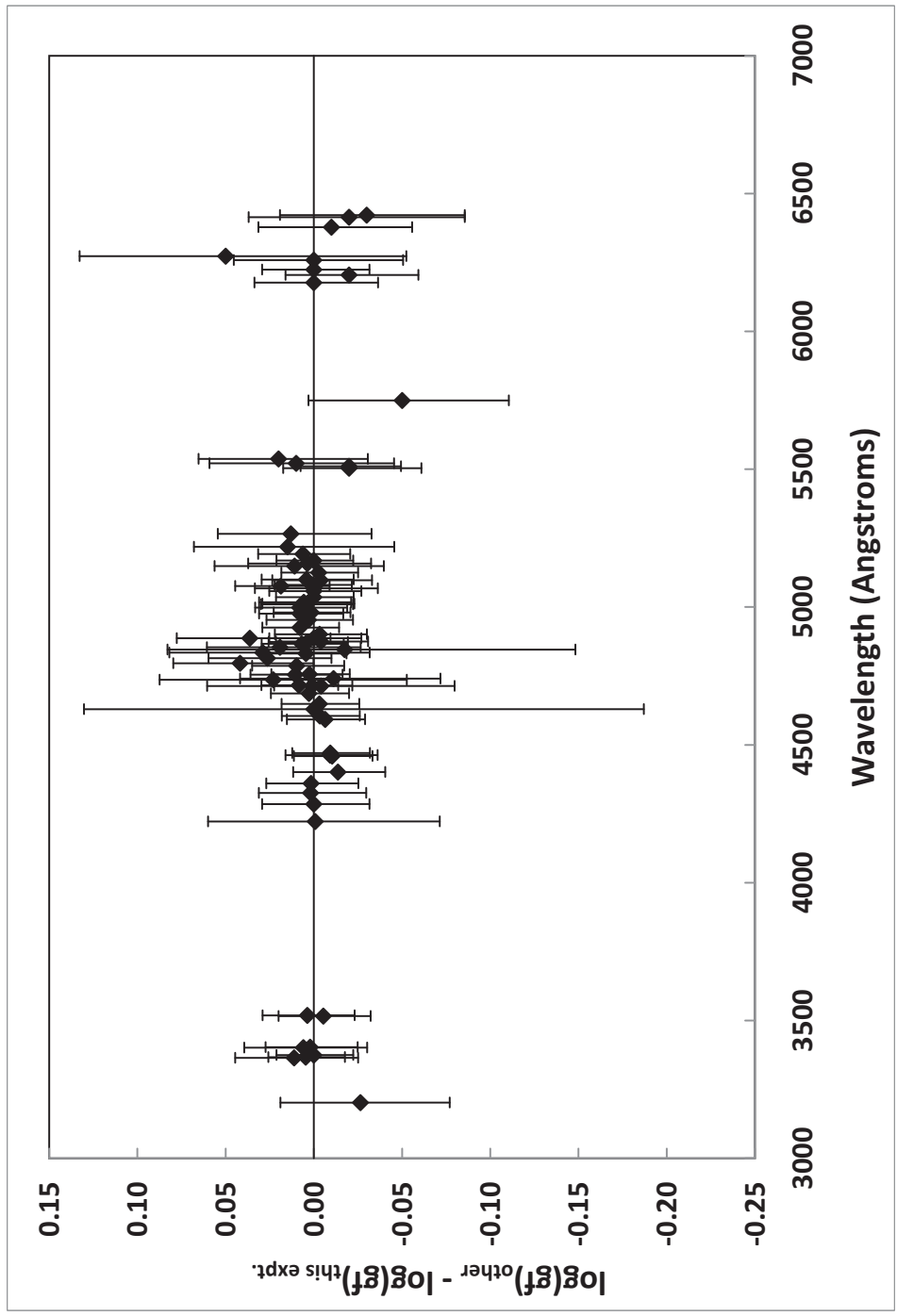

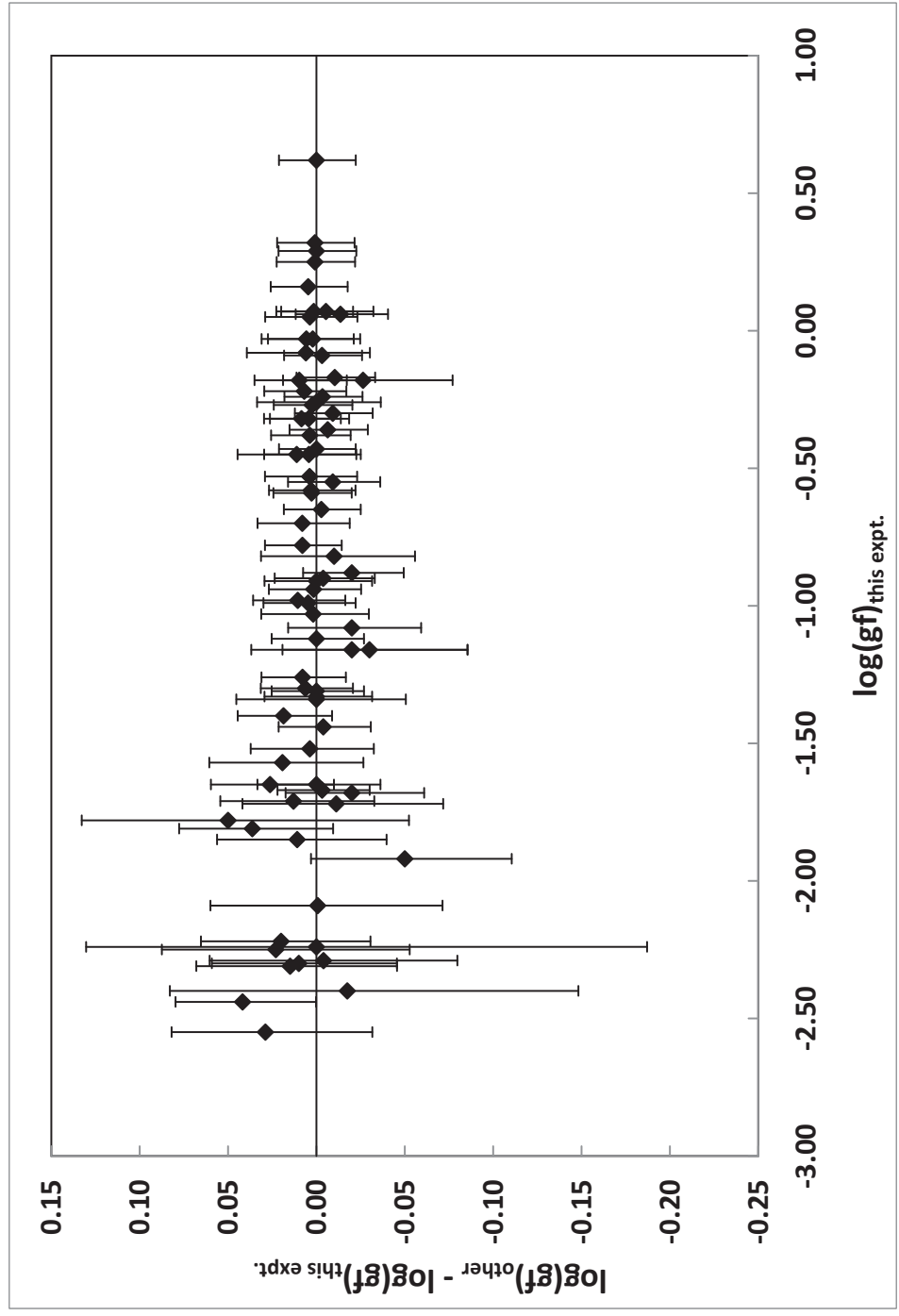

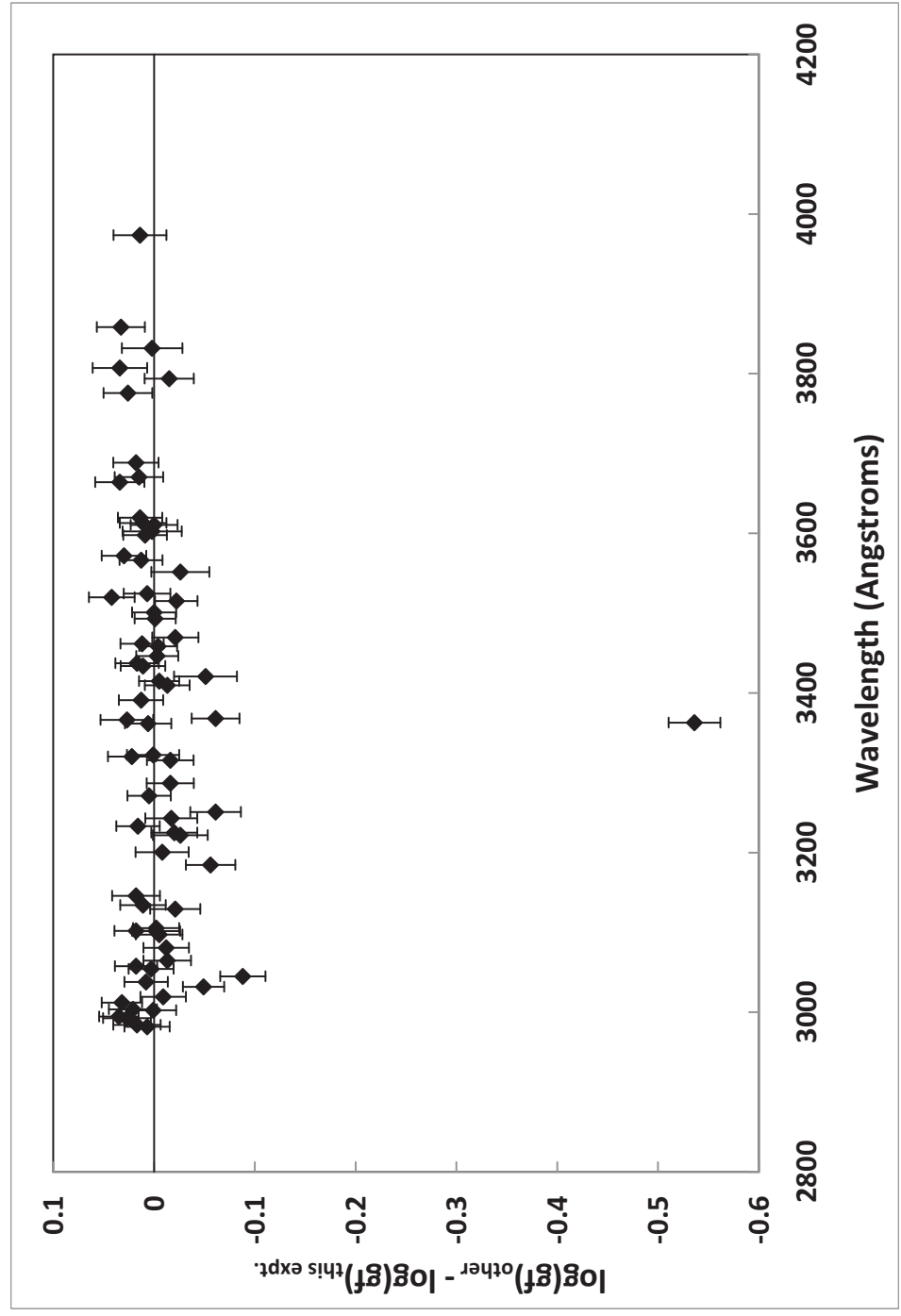

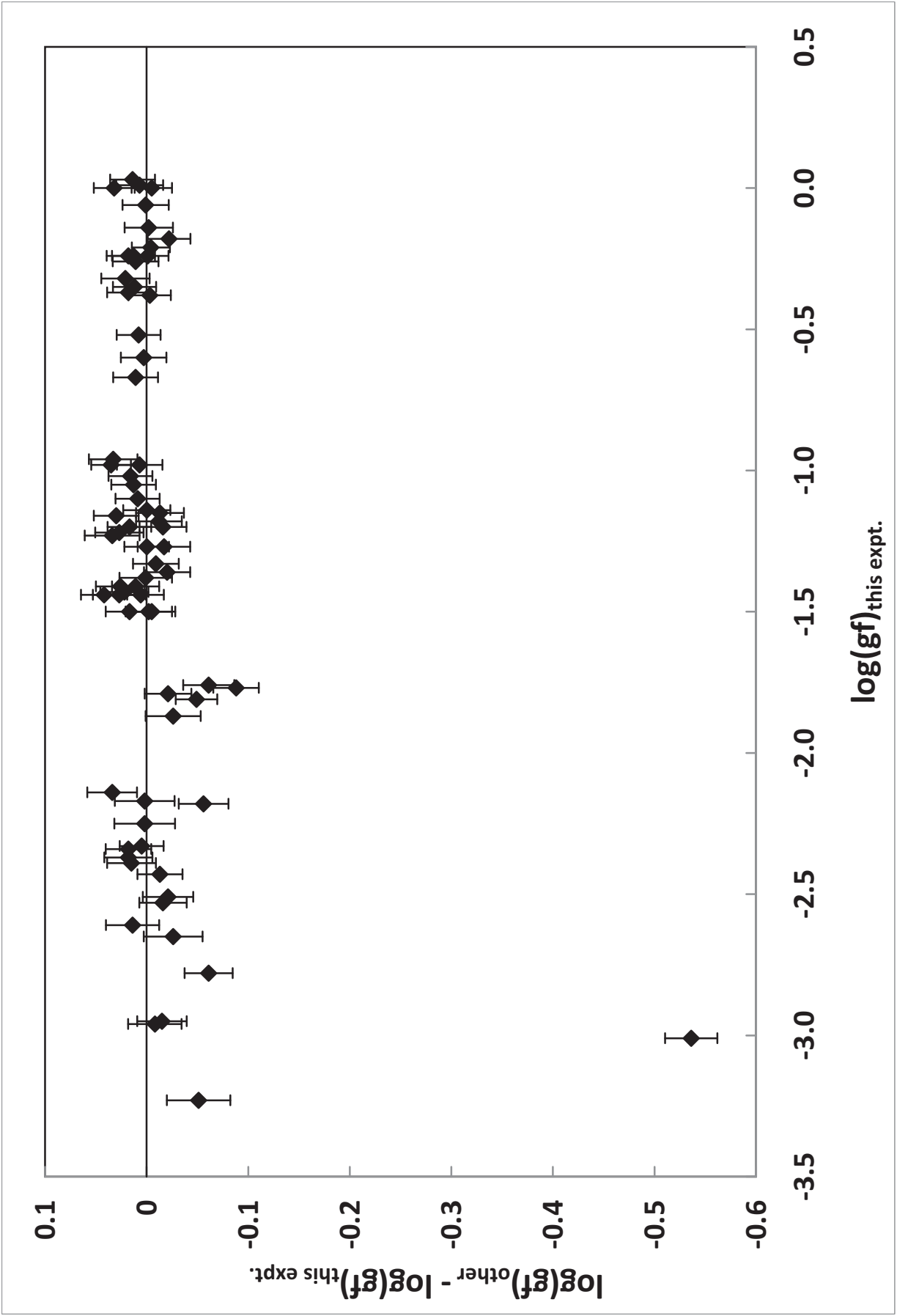

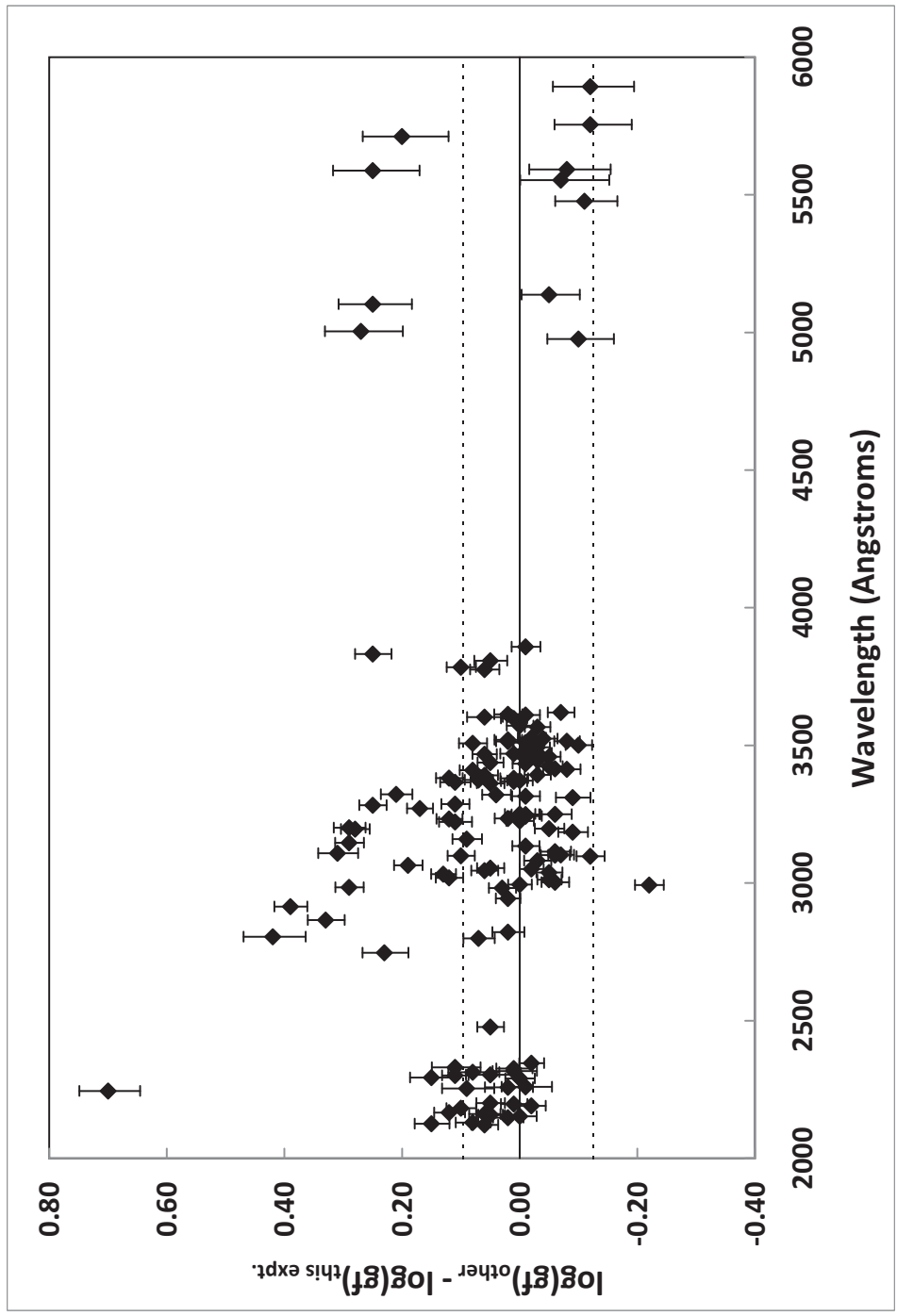

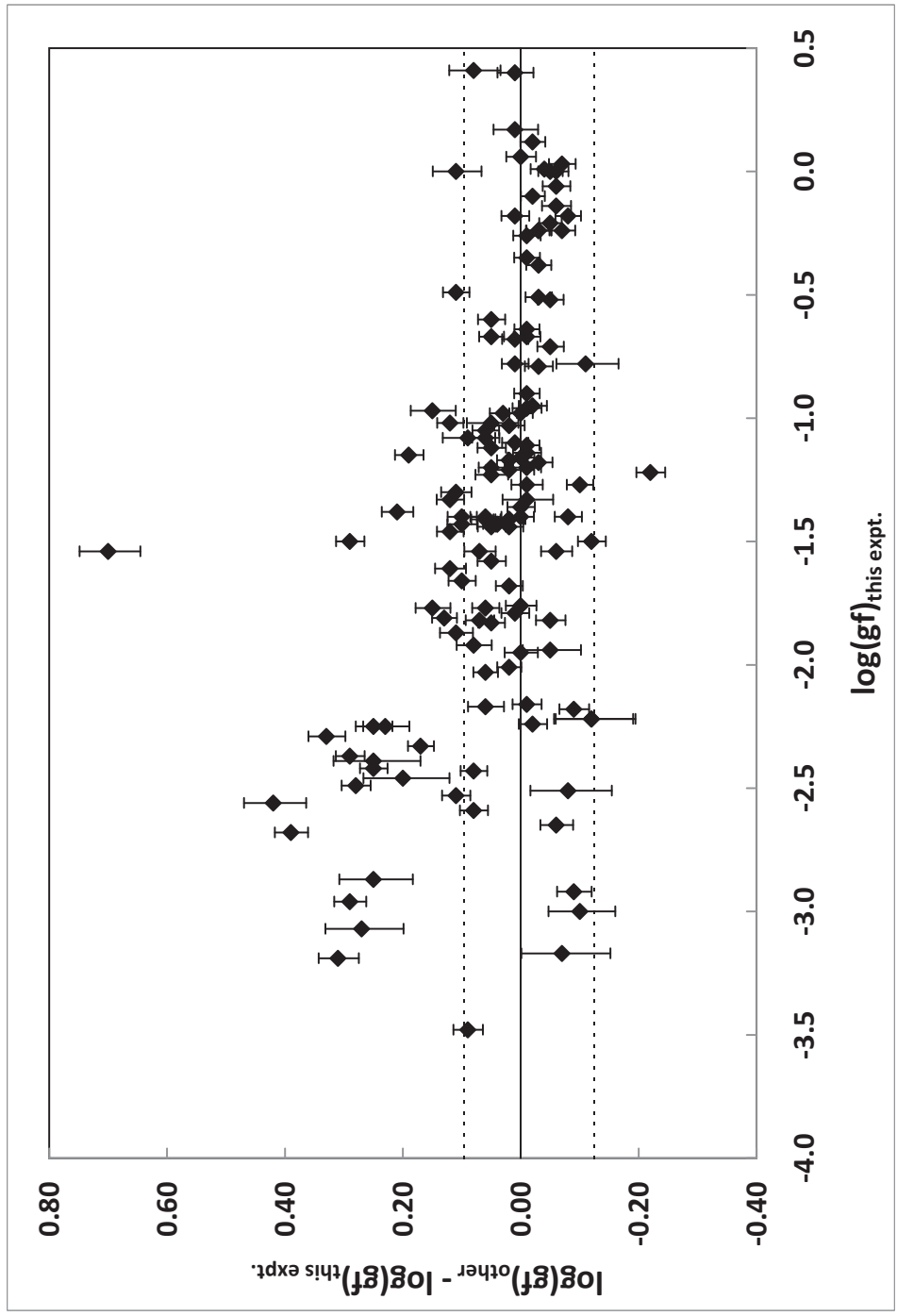

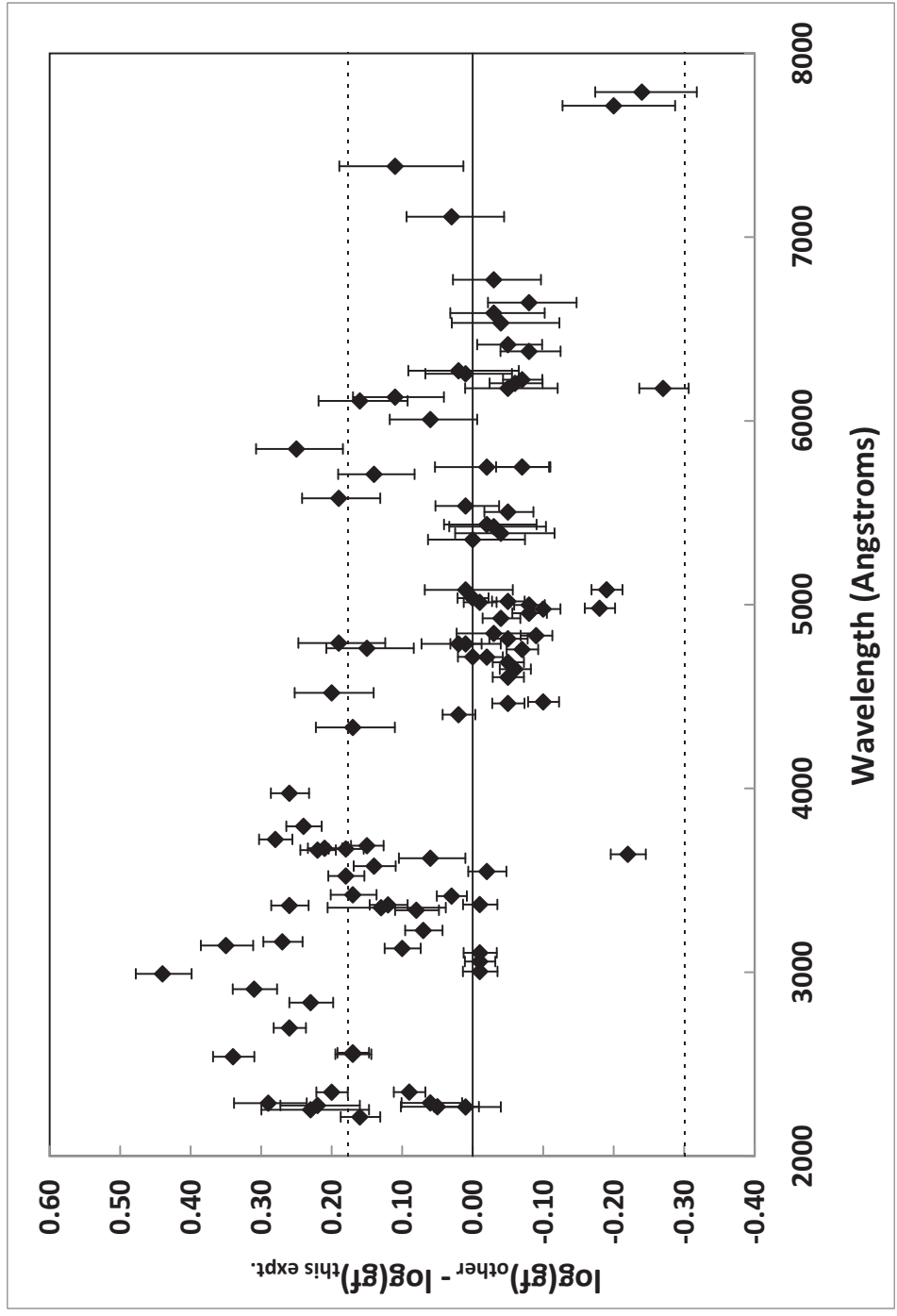

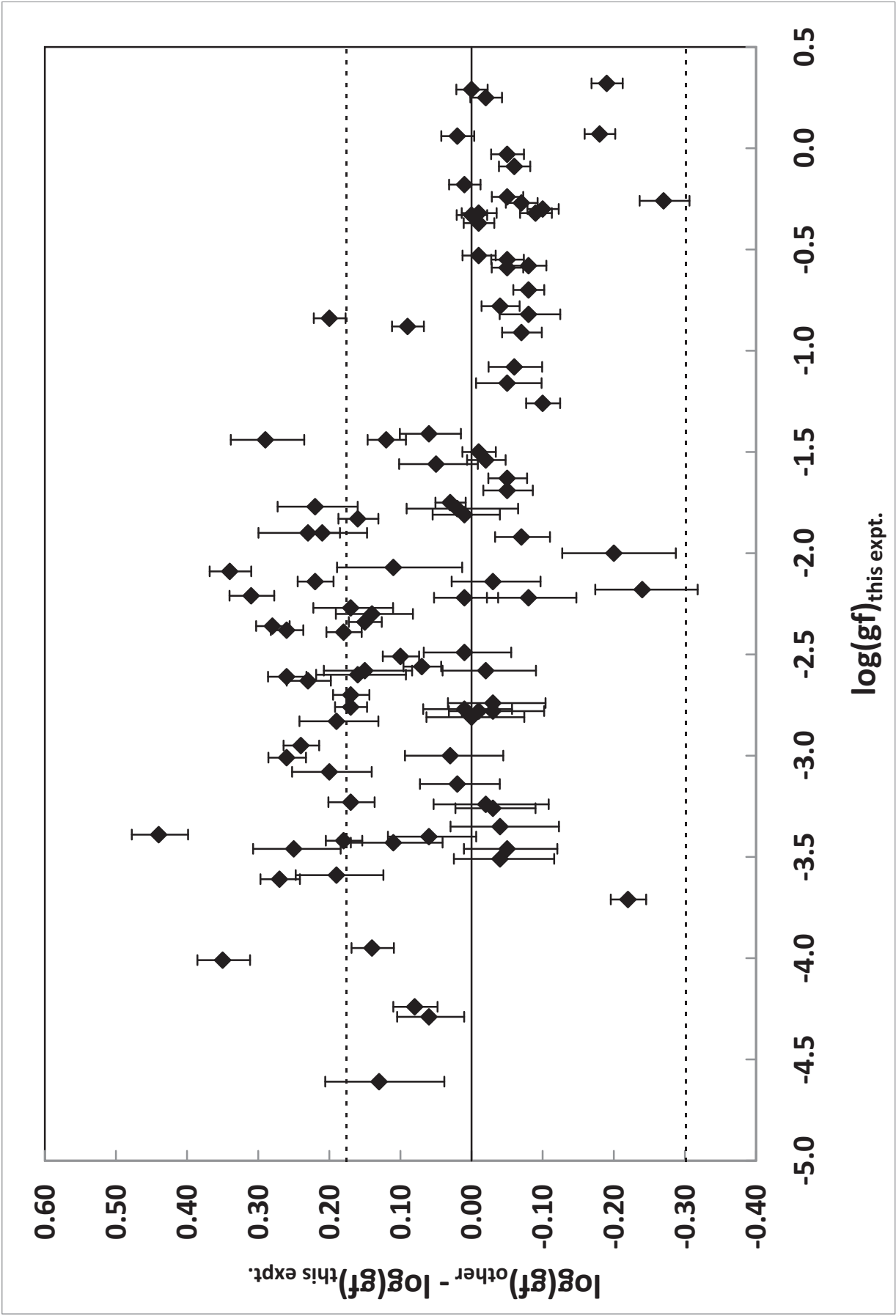

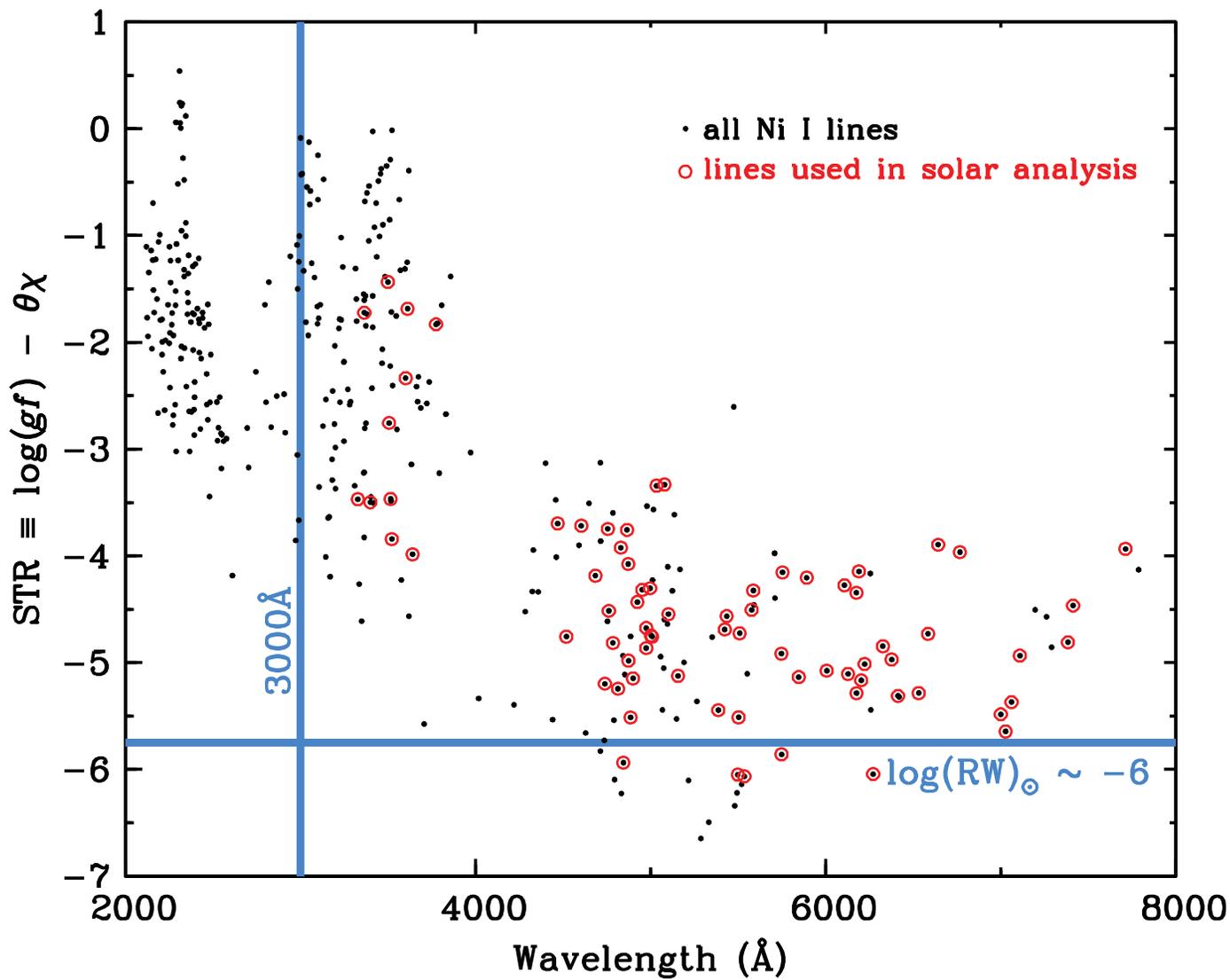

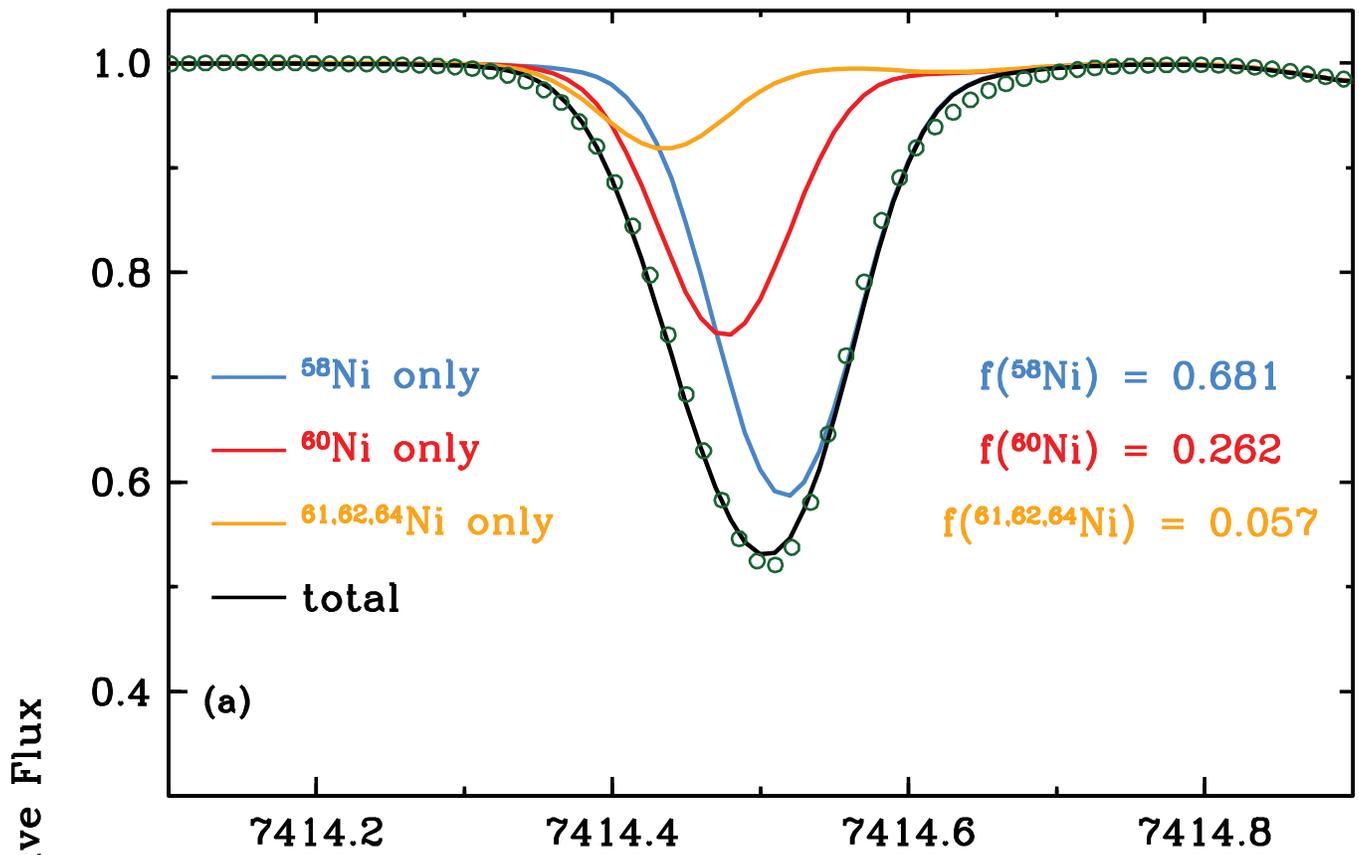
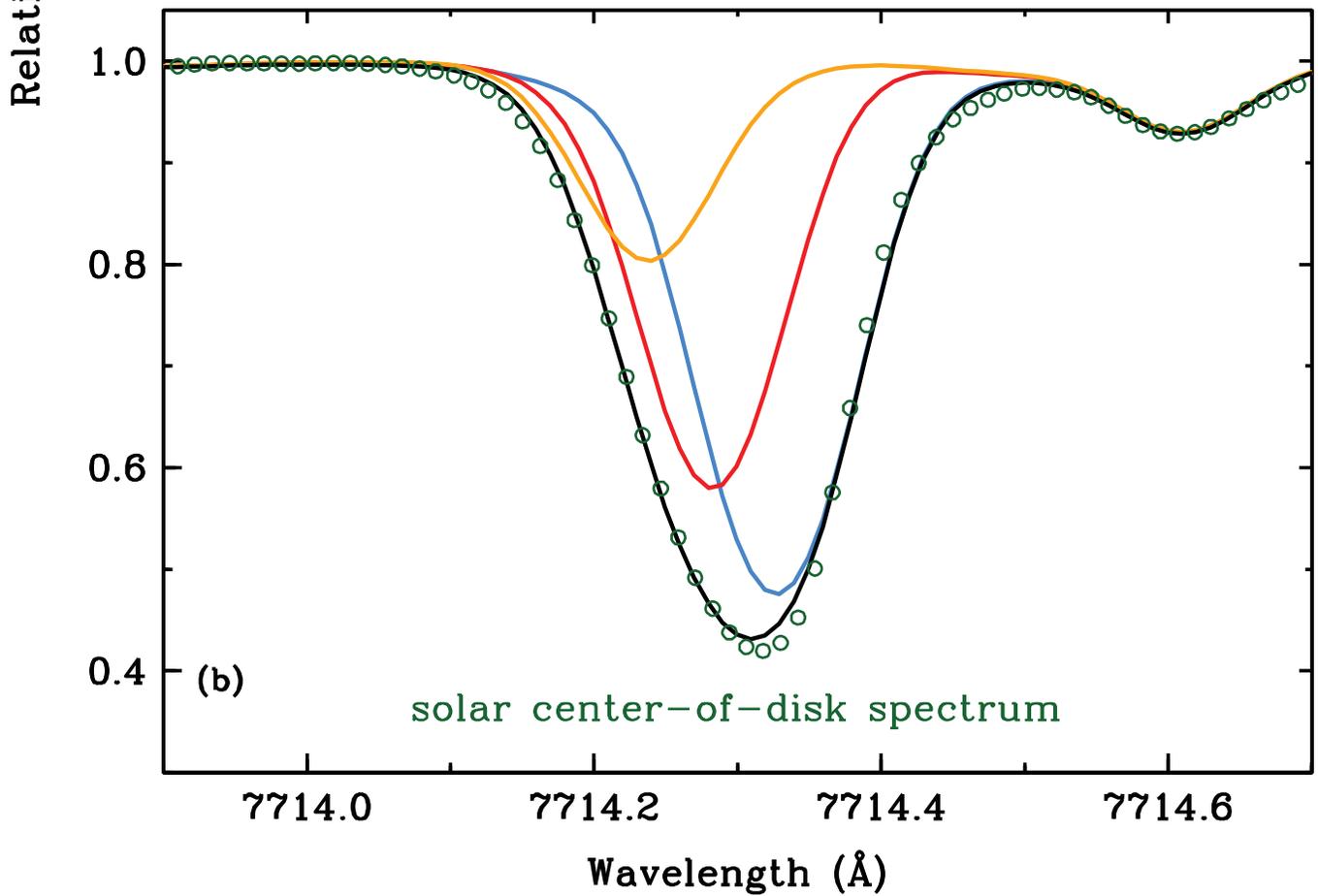

Table 4. Solar Photospheric Nickel Abundances from Individual Ni I Lines

| Wavelength in air (Å) | Lower Energy (eV) | $\log_{10}(gf)$ | $\log_{10}(\varepsilon)$ | Isotope[1] |
|---|---|---|---|---|
| 3328.713 | 0.109 | -3.36 | 6.25 | no |
| 3365.765 | 0.422 | -1.30 | 6.28 | no |
| 3401.164 | 3.417 | -0.08 | 6.15 | no |
| 3500.851 | 0.165 | -1.27 | 6.23 | no |
| 3507.693 | 0.165 | -2.59 | 6.25 | no |

Notes –Table 4 is available in its entirety via the link to the machine-readable version above.

[1]This column indicates whether isotopic substructure has been included in the synthetic spectrum of a line.

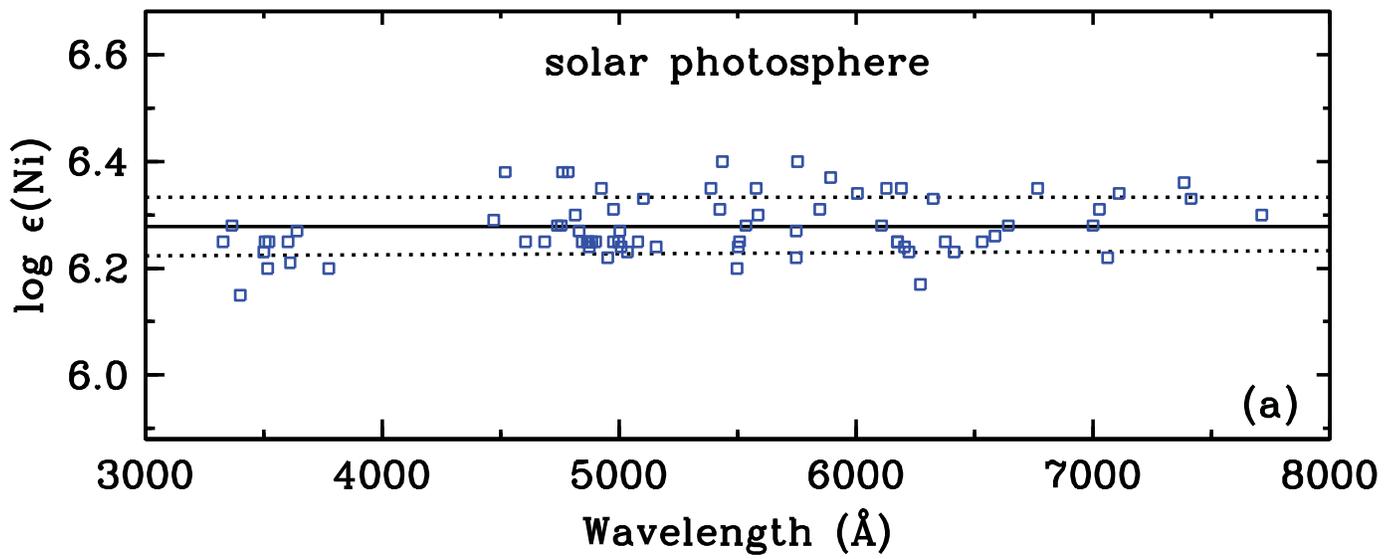
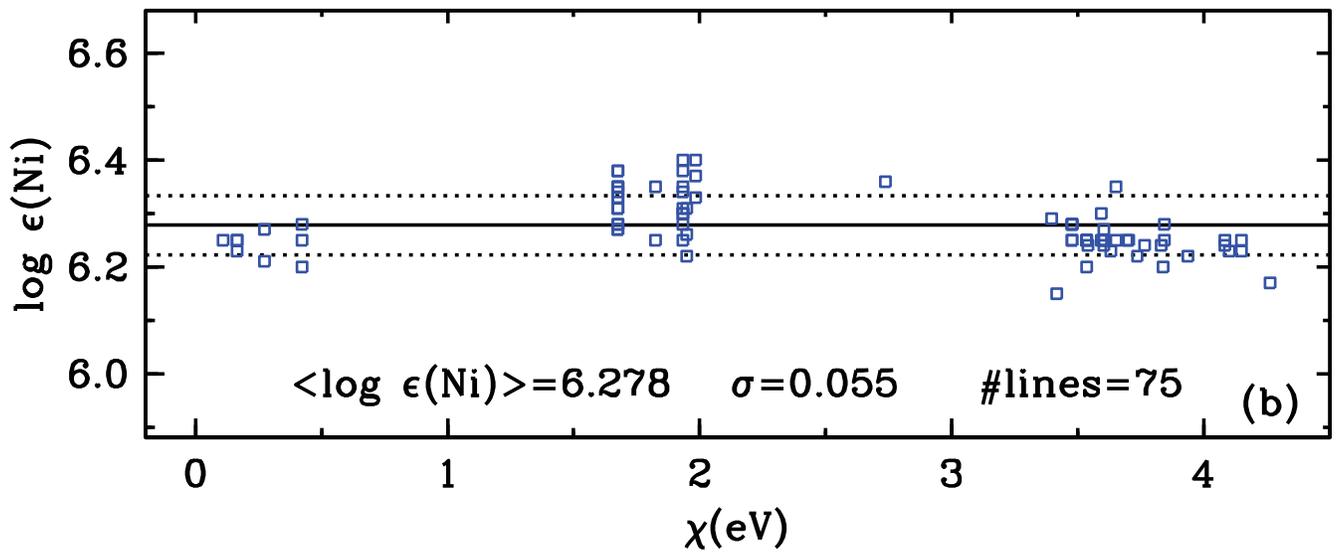
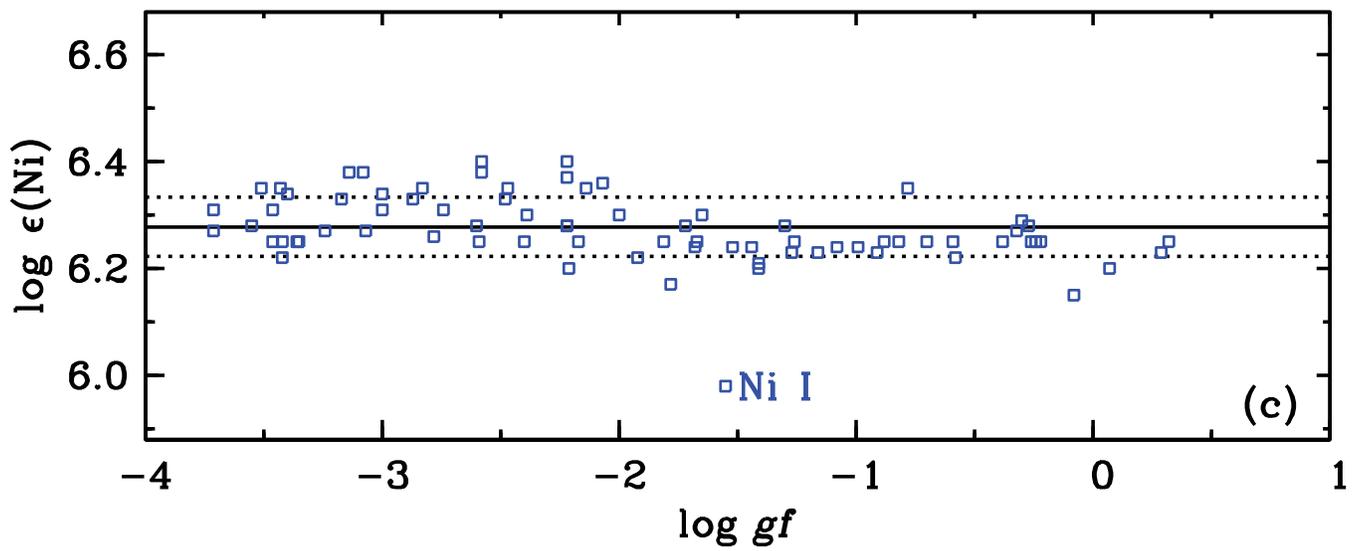

Table 5. Nickel Abundances from Individual Ni I Lines in HD 84937

| Wavelength in air (Å) | Lower Energy (eV) | $\log_{10}(gf)$ | $\log_{10}(\varepsilon)$ |
|---|---|---|---|
| 2318.777 | 0.422 | -1.73 | 3.87 |
| 2329.971 | 0.275 | +0.00 | 3.75 |
| 2338.496 | 0.025 | -2.03 | 4.00 |
| 2376.018 | 0.109 | -1.70 | 3.93 |
| 2401.842 | 0.165 | -1.10 | 3.90 |

Note –Table 5 is available in its entirety via the link to the machine-readable version above.

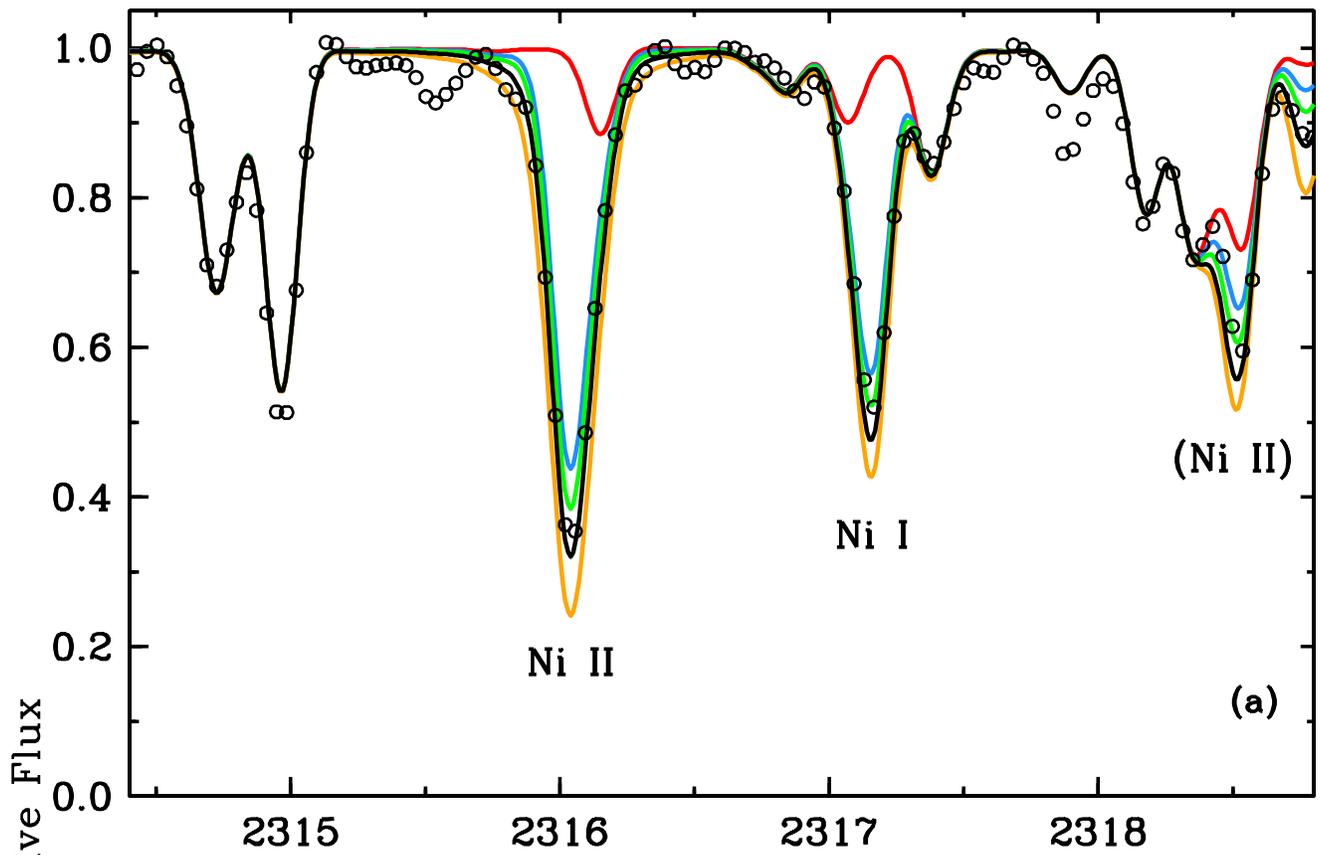
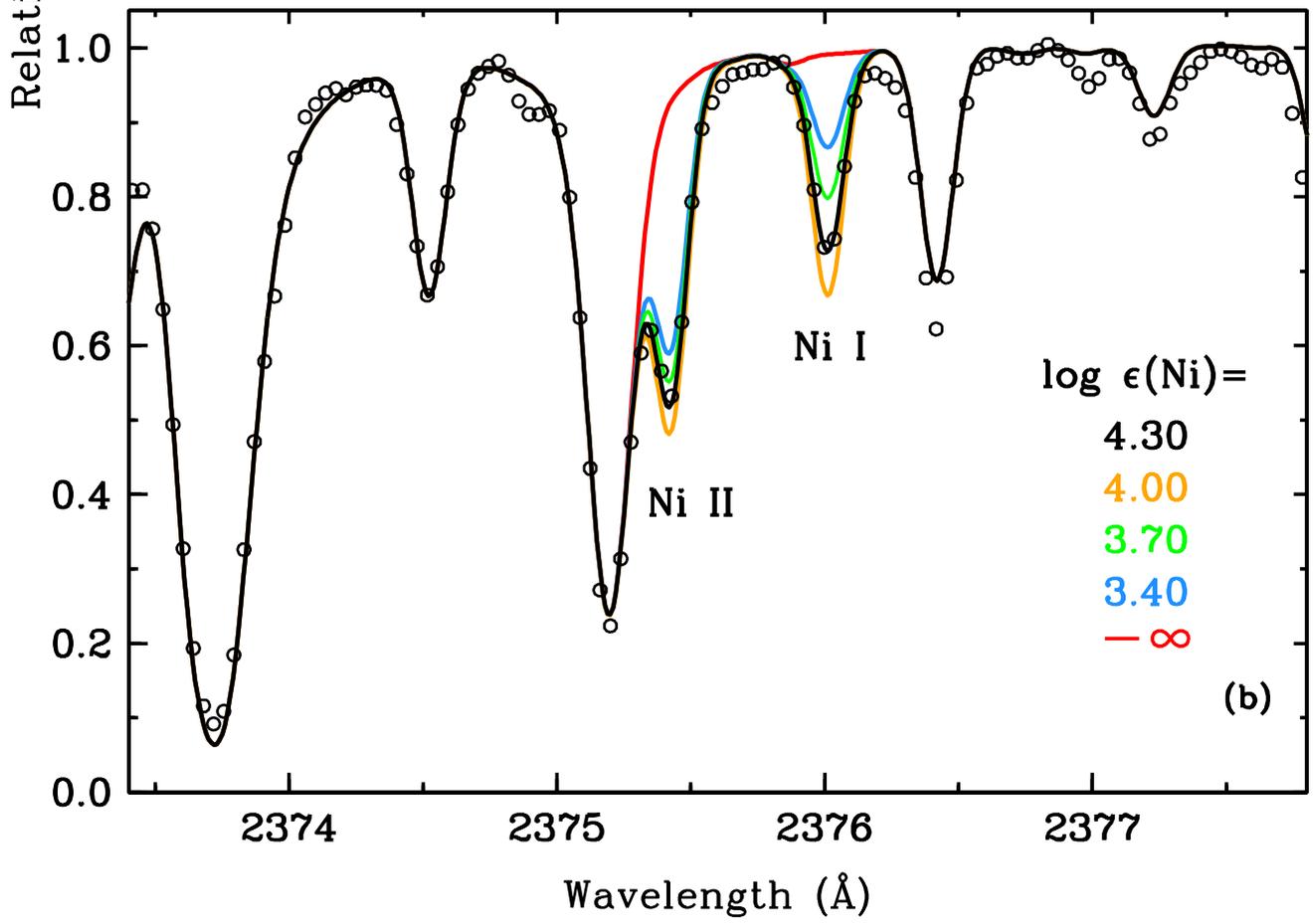

Table 6. Nickel Abundances from Individual Ni II Lines in HD 84937

| Wavelength in air (Å) | Lower Level eV | $\log_{10}(gf)$ | $\log_{10}(\varepsilon)$ |
|---|---|---|---|
| 2297.49 | 1.321 | -0.33 | 3.90 |
| 2316.04 | 1.040 | 0.27 | 3.92 |
| 2350.85 | 1.679 | -2.28 | 3.76 |
| 2367.38 | 1.156 | -1.29 | 3.80 |
| 2375.42 | 1.858 | -0.36 | 3.90 |
| 2387.76 | 1.679 | -1.07 | 4.10 |
| 2416.14 | 1.858 | 0.13 | 3.83 |
| 2437.89 | 1.679 | -0.33 | 3.90 |

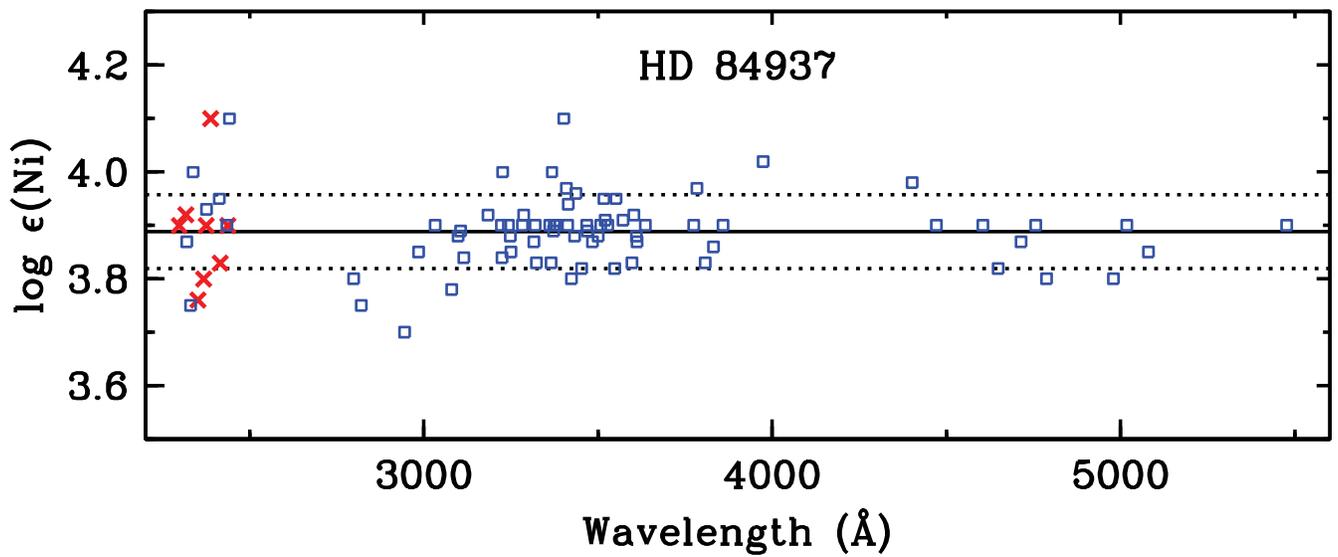
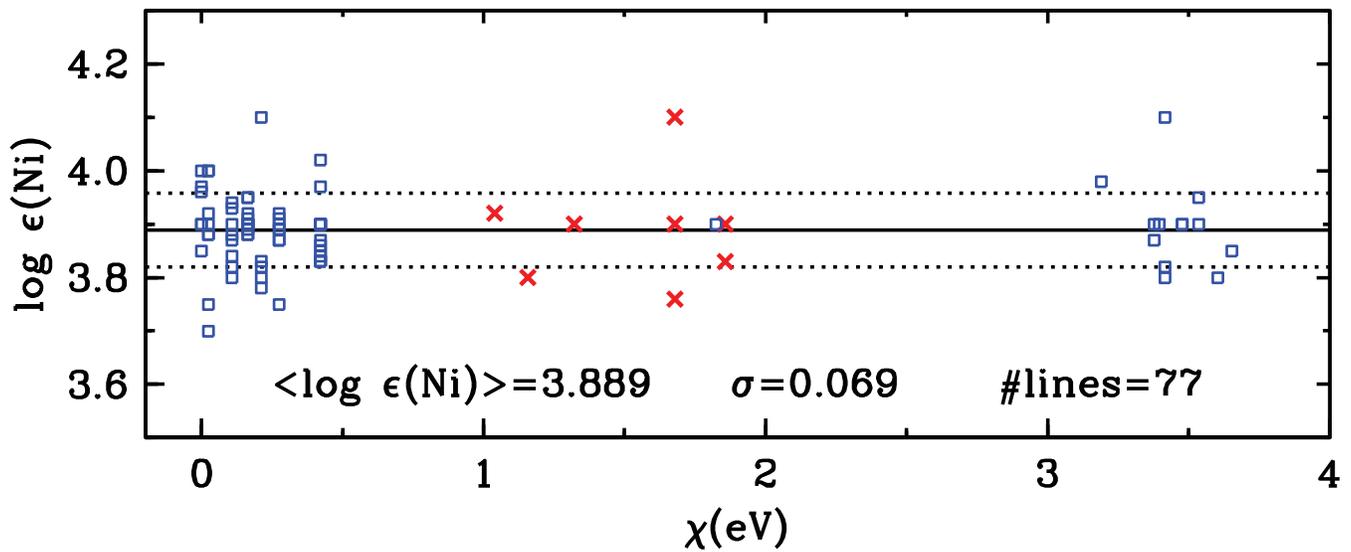
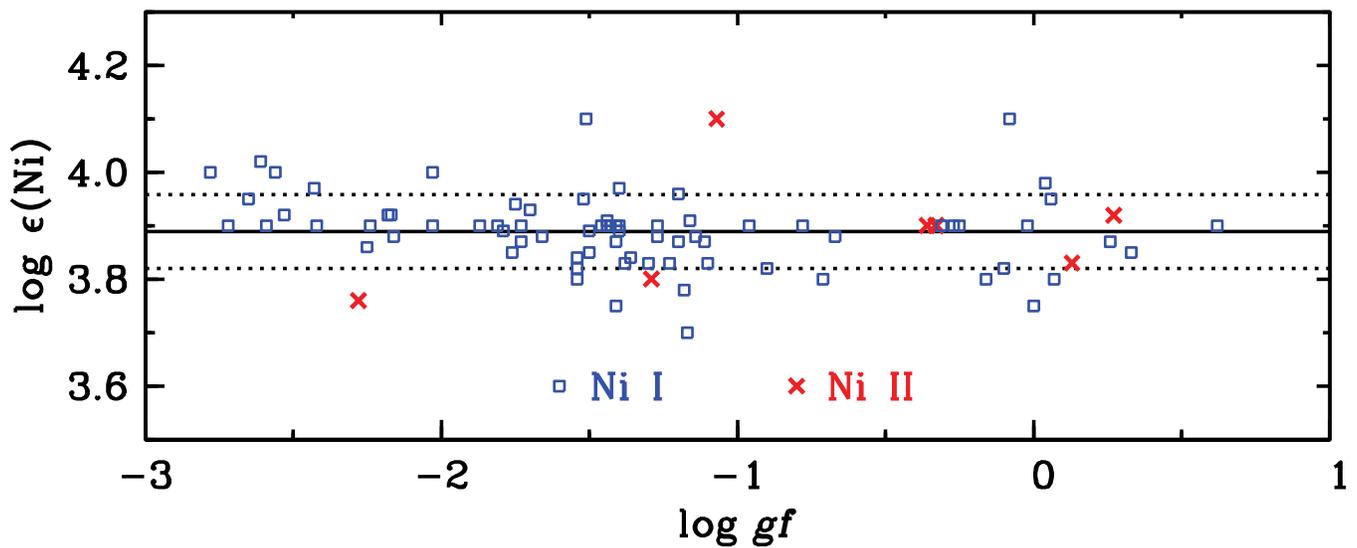

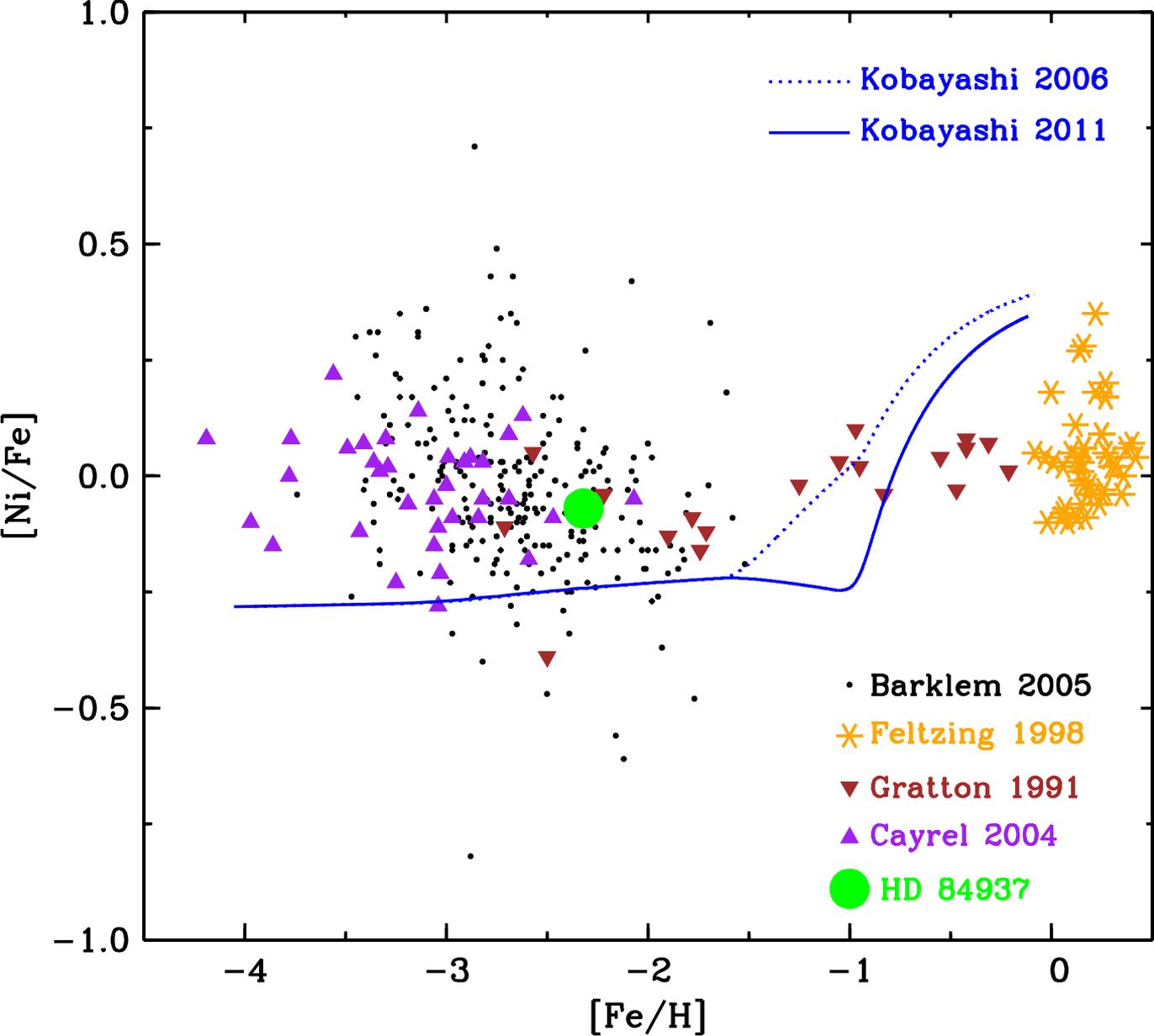

Table 7. Shifted isotopic wavelengths for 303 lines of Ni I organized by increasing center of gravity wavelength in air.

| Center of Gravity Wavelength in air[a] (Å) | Upper Level | | | Lower Level | | | Shifted $^{58}$Ni Wavelength in air[a] (Å) | Shifted $^{60}$Ni Wavelength in air[a] (Å) |
|---|---|---|---|---|---|---|---|---|
| | Energy[b] (cm$^{-1}$) | Parity | J | Energy[b] (cm$^{-1}$) | Parity | J | | |
| 2121.3903 | 47328.784 | od | 2 | 204.787 | ev | 3 | 2121.3898 | 2121.3910 |
| 2125.6261 | 47030.102 | od | 3 | 0.000 | ev | 4 | 2125.6265 | 2125.6254 |
| 2129.9541 | 47139.337 | od | 2 | 204.787 | ev | 3 | 2129.9535 | 2129.9550 |
| 2134.9235 | 47030.102 | od | 3 | 204.787 | ev | 3 | 2134.9228 | 2134.9244 |
| 2147.7838 | 47424.785 | od | 1 | 879.816 | ev | 2 | 2147.7834 | 2147.7845 |

Notes. –Table 7 is available in its entirety via the link to the machine-readable version above.

[a]Wavelength values computed from energy levels using the standard index of air from Peck & Reeder 1972.

[b]Energy levels from the online 2012 NIST Atomic Spectra Database by Kramida et al.